\documentclass[aps,reprint,superscriptaddress,longbibliography,nofootinbib]{revtex4-2}
\pdfoutput=1
\usepackage[utf8]{inputenc}

\usepackage{amsmath,amsthm,amssymb,amsfonts}
\usepackage{scalerel}
\usepackage{mathtools}
\usepackage{graphicx}
\usepackage{stfloats}
\usepackage[normalem]{ulem}
\usepackage[colorlinks = true,
            linkcolor = red,
            urlcolor  = titleblue,
            citecolor = cgreen,
            anchorcolor = blue]{hyperref}
\usepackage{mathrsfs}
\usepackage{bbold}
\usepackage{units}
\allowdisplaybreaks
\usepackage{bm}
\usepackage{braket}
\usepackage{makecell}
\usepackage[nameinlink]{cleveref} 
\usepackage{upgreek}
\usepackage{verbatim}
\usepackage{algorithm}
\usepackage[noend]{algpseudocode}
\usepackage[dvipsnames]{xcolor}
\usepackage{bbm}
\usepackage{array}
\usepackage{multirow}
\usepackage{tabularx}
\usepackage{float}
\usepackage{dcolumn}
\usepackage{slashed}
\usepackage{resizegather}
\usepackage[T1]{fontenc}
\usepackage[export]{adjustbox}

\theoremstyle{definition}

\theoremstyle{plain}

\theoremstyle{plain}
\newtheorem*{prop*}{Property}

\newtheorem*{ctr*}{Conjecture}

\definecolor{titleblue}{RGB}{37,96,167}
\definecolor{cgreen}{RGB}{11,200,10}

\begin{document}

\title{\textcolor{titleblue}{Full Eigenstate Thermalization in all-to-all Majorana Models}\\
}

\author{Abhik Kumar Saha\,\,\href{https://orcid.org/0000-0001-8168-6742}
{\includegraphics[scale=0.05]{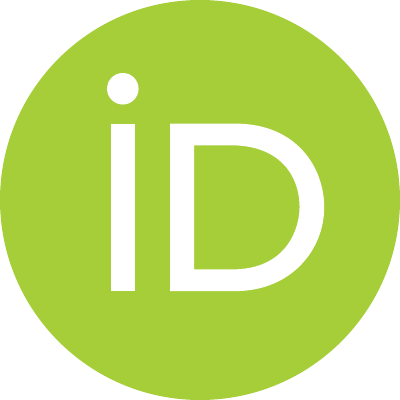}}}
\email{saha.abhikkumar.3k@kyoto-u.ac.jp}
\affiliation{Department of Physics, Kyoto University, Kitashirakawa Oiwakecho, Sakyo-ku, Kyoto 606-8502, Japan}

\begin{abstract}
The eigenstate thermalization hypothesis (ETH) provides a framework for understanding thermalization in isolated quantum many-body systems. Its generalization to higher-order correlations of matrix elements, known as full ETH, establishes a connection with free probability through ETH free cumulants. In this work, we numerically investigate full ETH in all-to-all Sachdev-Ye-Kitaev (SYK) Majorana systems using exact diagonalization, comparing chaotic quartic $\rm SYK_{4}$, integrable quadratic $\rm SYK_{2}$, and integrable disorder-free quartic $\rm SYK_{4}$ models. We test the validity of the approximate decomposition of the out-of-time-ordered correlator into contributions involving second and fourth order ETH free cumulant in all three models. We further examine the finite-size corrections to the full-ETH conditions and find substantially slower, operator-dependent suppression in the integrable models than in the chaotic case. Remarkably, both integrable models exhibit similar finite-size scaling despite their different interaction structures.
\end{abstract}

\maketitle

\emph{Introduction.} How thermal behavior emerges from the unitary dynamics of isolated quantum many-body systems is a central question in quantum statistical mechanics. The eigenstate thermalization hypothesis (ETH) provides a microscopic understanding of this phenomenon through the statistical properties of the matrix elements of observables in the energy eigenbasis. According to standard ETH, the diagonal matrix elements of a local observable vary smoothly with energy, while the off-diagonal matrix elements exhibit pseudorandom fluctuations whose typical magnitude is exponentially suppressed with increasing system size. The validity of ETH has been extensively investigated in nonintegrable quantum many-body systems \cite{PhysRevE.60.3949,Rigol_2008,PhysRevLett.105.250401,RevModPhys.83.863,PhysRevE.87.012125,D_Alessio_2016,PhysRevLett.117.170404,PhysRevB.93.134201,PhysRevLett.129.030602,PhysRevA.43.2046,PhysRevE.50.888,deutsch2025closedquantumgivingergodicity}. However, standard ETH does not fully account for correlations among multiple matrix elements. Such correlations are particularly relevant for understanding quantum chaos and information scrambling via the out-of-time-ordered correlator (OTOC), and have attracted considerable attention in recent studies \cite{PhysRevE.99.042139,PhysRevLett.123.260601,PhysRevLett.122.220601,PhysRevLett.123.230606,PhysRevE.104.034120,PhysRevE.102.042127,Hashimoto_2017,Roberts_2017,Hosur_2016,Maldacena_2016}. 

The generalization of standard ETH, known as \textit{full ETH} \cite{PhysRevE.99.042139,PhysRevLett.129.170603,PhysRevLett.134.140404,d1cq-gj9p,Vallini_2026,PhysRevB.111.054303,6n9v-b3b7}, incorporates higher-order correlations among matrix elements and establishes a connection with free probability through ETH free cumulants. Numerical investigations of full ETH have primarily focused on chaotic and integrable quantum many-body systems with \emph{local} interactions \cite{PhysRevLett.134.140404,n7yh-828k,zhang2026finitesizescalingeigenstatethermalization}. This naturally raises the question: \textit{What is the fate of the full ETH in integrable and chaotic many-body systems with all-to-all interactions?} To answer this question, we turn to the Sachdev–Ye–Kitaev ($\rm SYK_{p}$) models, a quintessential class of all-to-all quantum many-body systems consisting of $N$ Majorana fermions with $p$-body interactions \cite{PhysRevLett.70.3339,kitaev1998,kitaev2015}. These models have attracted considerable attention across various fields such as quantum chaos, quantum information, and high-energy physics to name a few \cite{PhysRevD.94.106002,PhysRevX.5.041025,PhysRevB.105.075117,PhysRevB.105.235131,PhysRevB.94.035135,PhysRevB.100.155128,Gross_2017,Li_2017,Fu_2017,PhysRevX.12.021040,Gu_2017,Berkooz_2018,PhysRevLett.105.151602,PhysRevD.94.106002,Cotler_2017,RevModPhys.94.035004,faulkner2022snowmasswhitepaperquantum}. Previous investigations of standard ETH in $\rm SYK_{p}$ models have established that chaotic quartic $\rm SYK_{4}$ exhibits the expected ETH scaling, whereas quadratic integrable $\rm SYK_{2}$ displays a much slower algebraic suppression of diagonal matrix-element fluctuations with system size \cite{PhysRevB.100.115122}. While the random-coupling quartic $\rm SYK_{4}$ model is chaotic, its disorder-free counterpart with uniform four-body couplings is integrable \cite{PhysRevResearch.7.013092}. Here, we investigate full ETH by numerically studying three all-to-all interacting many-body models: chaotic quartic $\rm SYK_{4}$, integrable quadratic $\rm SYK_{2}$ \cite{PhysRevLett.120.241603,PhysRevLett.125.250602}, and integrable disorder-free quartic $\rm SYK_{4}$ \cite{PhysRevResearch.7.013092}. Specifically, we study the approximate decomposition of the OTOC into contributions involving second and fourth-order ETH free cumulants and examine the finite-size scaling of crossing contributions and deviations from cactus factorization for bilinear and quartic Majorana observables. We find substantially slower, observable-dependent suppression of these finite-size corrections in the integrable models compared with the chaotic case. Notably, both integrable models exhibit similar effective scaling behavior despite their different interaction structures.

\emph{Full ETH.} The standard ETH \cite{PhysRevA.43.2046,PhysRevE.50.888,deutsch2025closedquantumgivingergodicity} characterizes the matrix elements of a local observable $O$ in the energy eigenbasis of a Hamiltonian $H$, with $H|i\rangle=E_{i}|i\rangle$. The ETH ansatz is written as
\begin{equation}
O_{ij}={O}(E_{i})\delta_{ij}+e^{-S(E^+)/2}
F^{(2)}_{E^+}(\omega_{ij})R_{ij},
\end{equation}
where $O_{ij}=\langle i|O|j\rangle$, $E^{+}=(E_{i}+E_{j})/2$, and $\omega_{ij}=E_{i}-E_j$. $O(E)$ is the thermal expectation value, $S(E^{+})$ is the thermodynamic entropy, $F^{(2)}_{E^+}(\omega)$ is a smooth function decaying at large $\omega$, and $R_{ij}$ is a pseudorandom variable with zero mean and unit variance. While standard ETH describes the statistical properties relevant to one and two-point functions \cite{D_Alessio_2016}, it does not fully characterize correlations among multiple matrix elements.

Full ETH extends this framework to higher-order correlations among matrix elements. For a cyclic product of $q$ matrix elements with mutually distinct energy indices, the full ETH ansatz is
\begin{equation}
\overline{O_{i_1i_2}O_{i_2i_3}\cdots O_{i_qi_1}}=e^{-(q-1)S(E^{+})}F_{e^{+}}^{(q)}(\omega).
\label{eq:full_ETH}
\end{equation}
Here, the \textit{overbar} on the left hand side denotes averaging, $e^{-S(E^{+})}$ is the mean level spacing at average energy $E^{+}=\frac{1}{q}\sum_{a=1}^{q}E_{i_a}$, $\omega=(\omega_{i_1i_2},\omega_{i_2i_3},\cdots,\omega_{{i}_{q-1}i_{q}})$, and $\omega_{ij}=E_{i}-E_j$ are the energy differences. $F_{e^{+}}^{(q)}(\omega)$ is a smooth function of the energy density $e^{+}=E^{+}/N$ and $\omega$. Eq.~\eqref{eq:full_ETH} is referred as \textit{full ETH}. These higher-order correlations are connected to multipoint thermal correlation functions through free probability \cite{PhysRevLett.129.170603}. In particular, a thermal $q$-point correlation function follows
\begin{equation}
\langle O(t_1)O(t_2)\cdots O(t_q)\rangle_{\beta}=\sum_{\pi\in NC(q)}\kappa_{\pi}^{\beta}(t_{1},t_{2},\cdots,t_{q}),
\label{eq:q_point}
\end{equation}
where $\langle \bullet\rangle_{\beta}=\frac{1}{Z}\mathrm{Tr}(e^{-\beta H}\bullet)$, and $Z=\mathrm{Tr}(e^{-\beta H})$. Here $NC(q)$ denotes the set of non-crossing partitions of $q$ elements, and $\kappa_{\pi}^{\beta}$ denotes the product of thermal free cumulants $k_{n}^{\beta}$ associated with the blocks of a particular non-crossing partition $\pi$. The full ETH ansatz given by Eq.~\eqref{eq:full_ETH} then relates these thermal free cumulants to quantities constructed directly from the matrix elements of $O$. More explicitly, the $q$-th order ETH free cumulant is defined as 
\begin{equation}
k_{q}^{\beta}({\textbf{t}})\simeq k_{q}^{\rm ETH}({\textbf{t}})=\frac{1}{Z}\sum_{i_1\neq i_2\neq\cdots i_q}e^{-\beta E_{i_1}}O(t_1)_{i_1i_2}O(t_2)_{i_2i_3}\cdots O(t_q)_{i_qi_1}. 
\label{Eq:k}
\end{equation}
Here, $O(t)_{ij}=e^{i(E_i-E_j)t}O_{ij}$, and the relation $k_{q}^{\beta}({\textbf{t}})\simeq k_{q}^{\rm ETH}({\textbf{t}})$ is expected to hold in the thermodynamic limit. Higher-order matrix element correlations can be represented diagrammatically through crossing and noncrossing contributions, with crossing terms first appearing at fourth order. Full ETH predicts that crossing contributions are subleading. In particular,
\begin{equation}
{\rm cross}(t)=\frac{1}{\mathcal D}\sum_{i\neq j}
e^{i\omega_{ij}t}|O_{ij}|^4
=\mathcal O(\mathcal D^{-a}),\ a>0.
\label{eq:cross}
\end{equation}
In contrast, noncrossing contributions associated with cactus diagrams are expected to factorize. Specifically,
\begin{equation}
\begin{aligned}
{\rm cac}(t_1,t_2)
&=\frac{1}{\mathcal D}
\sum_{i\neq j\neq k}
e^{i\omega_{ij}t_1+i\omega_{jk}t_2}
|O_{ij}|^2|O_{jk}|^2\\
&\approx k_2^{\rm ETH}(t_1)k_2^{\rm ETH}(t_2).
\end{aligned}
\label{eq:cac}
\end{equation}
Here, $\mathcal{D}$ is the Hilbert-space dimension, and $k_2^{\rm ETH}(t)$ is given by Eq.~\eqref{Eq:k}. This factorization is expected in the thermodynamic limit for generic times $t_1$ and $t_2$. The suppression of crossing contributions and deviations from cactus factorization therefore provide complementary probes of the finite-size behavior of the full-ETH conditions \cite{PhysRevLett.129.170603,PhysRevLett.134.140404}. The conditions given by Eqs.~\eqref{eq:cross} and \eqref{eq:cac} hold both in time as well as frequency domain.


\emph{Models.} We investigate three all-to-all Majorana models to compare the full-ETH conditions in chaotic and integrable systems, including both noninteracting and interacting integrable cases. We begin with the chaotic quartic $\rm SYK_{4}$ model \cite{PhysRevD.94.106002}, whose Hamiltonian is given by
\begin{equation}
H_{\rm SYK_4}^{\rm (C)}=\sqrt\frac{6}{N^{3}}\sum_{1\leq i<j<k<l\leq N} J_{ijkl}\chi_{i}\chi_{j}\chi_{k}\chi_{l},
\label{eq:SYK4C}
\end{equation}
where $\chi_{i}$ are the Majorana fermion operators which satisfy the Clifford algebra: $\{\chi_{i},\chi_{j}\}=\delta_{ij}$ and $J_{ijkl}$ are independent Gaussian random variables with zero mean and unit variance with probability distribution function $P(J_{ijkl})=\frac{1}{\sqrt{2\pi}}e^{-J_{ijkl}^2/2}$. 

We next consider the quadratic $\rm SYK_2$ model \cite{PhysRevLett.120.241603,PhysRevLett.125.250602},
\begin{equation}
H_{\rm SYK_2}^{\rm (I)}=i\sqrt\frac{1}{N}\sum_{1\leq i<j\leq N} J_{ij}\chi_{i}\chi_{j},
\label{eq:SYK2I}
\end{equation}
where $J_{ij}$ are independent Gaussian random variables with zero mean and unit variance. Being quadratic in Majorana operators, this model is noninteracting and integrable. To investigate an interacting all-to-all integrable system, we further consider the disorder-free quartic $\rm SYK_4$ model with uniform couplings, $J_{ijkl}=1$ \cite{PhysRevResearch.7.013092}, 
\begin{equation}
H_{\rm SYK_4}^{\rm (I)}=\sqrt\frac{6}{N^{3}}\sum_{1\leq i<j<k<l\leq N}\chi_{i}\chi_{j}\chi_{k}\chi_{l},
\label{eq:SYK4I}
\end{equation}
which contains four-Majorana interactions but remains integrable due to its uniform coupling structure. To investigate the full ETH conditions, we consider bilinear and quartic Majorana observables, $\hat A_{ij}=i\chi_{i}\chi_{j}$ and $\hat B_{ijkl}=\chi_{i}\chi_{j}\chi_{k}\chi_{l}$, with mutually distinct Majorana indices. The symmetry properties of the $\rm SYK_{4}$ model exhibit a periodic dependence of the system size $N$, corresponding to different random-matrix ensembles \cite{PhysRevB.95.115150,PhysRevD.94.126010}. To ensure a consistent comparison across system sizes, we consider $N$ within a common ensemble. In the Gaussian orthogonal ensemble (GOE) class, the diagonal matrix elements of $\hat A_{ij}$ vanish identically for all energy eigenstates \cite{PhysRevB.100.115122}, while the Gaussian symplectic ensemble (GSE) class contains additional symmetry sectors within a fixed fermion-parity sector. Moreover, the Gaussian unitary ensemble (GUE) class provides more numerically accessible system sizes than either of the other two. We therefore focus on systems with GUE symmetry, corresponding to the sequence $N=2+4m$, with $m\in\mathbb{Z}$. All three Hamiltonians conserve fermion parity, and we restrict our analysis to the even-parity sector with Hilbert-space dimension $\mathcal{D}=2^{N/2-1}$. We perform full exact diagonalization for system sizes up to $N=30$. For the two disordered Hamiltonians, $H_{\rm SYK_4}^{\rm C}$ and $H_{\rm SYK_2}^{\rm I}$, we average over $2^{20-N/2}$ disorder realizations for each system size. 

\begin{figure}[h!]
    \centering
    \includegraphics[width=0.8\columnwidth]{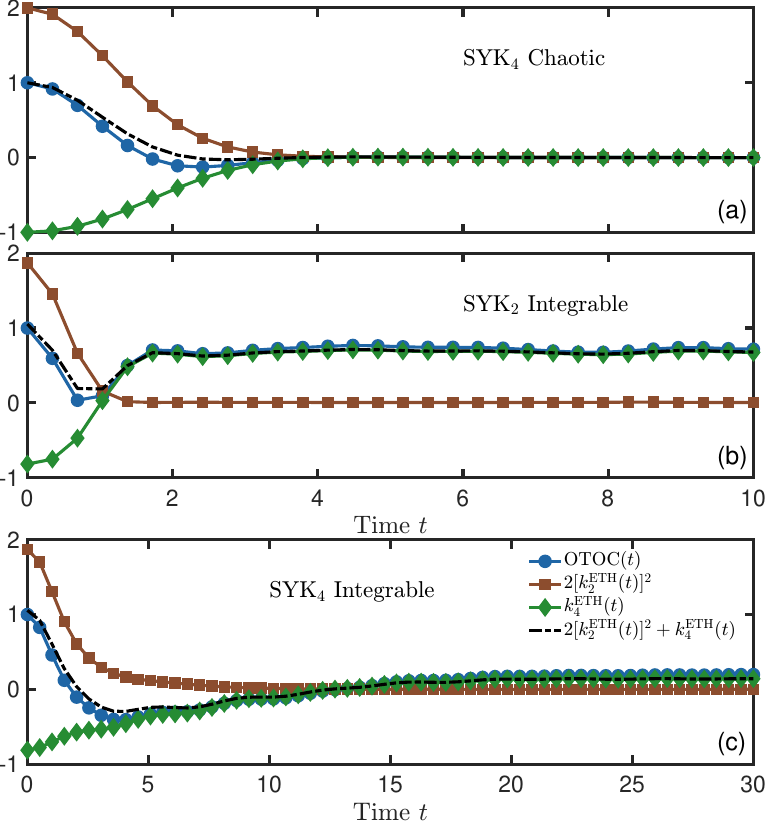}
    \caption{The time evolution ${\rm OTOC}(t)$ compared with $2[k_{2}^{\rm ETH}(t)]^2$ and $k_4^{\rm ETH}(t)$ and their sum for (a) chaotic quartic $\rm SYK_4$, (b) integrable quadratic  $\rm SYK_2$, and (c) integrable disorder-free quartic $\rm SYK_4$. Results are shown for the bilinear Majorana observable $\hat A=i\chi_{1}\chi_{2}$ with $N=30$, averaged over $32$ disorder realizations for the disordered models.}
    \label{fig:all_model_OTOC}
\end{figure}

\emph{Results.} We first examine the relation between the OTOC and the ETH free cumulants at infinite temperature ($\beta=0$). For a traceless observable, the full ETH predicts \cite{PhysRevE.99.042139,PhysRevLett.129.170603,PhysRevLett.134.140404}
\begin{equation}
{\rm OTOC}(t)\simeq 2[k_2^{\rm ETH}(t)]^2+k_4^{\rm ETH}(t),
\label{eq:OTOC}
\end{equation}
where ${\rm OTOC}(t)=\langle O(t)O(0)O(t)O(0)\rangle_{\beta=0}$ follows from Eq.~\eqref{eq:q_point}, $k_2^{\rm ETH}(t)$ and $k_4^{\rm ETH}(t)$ are the second and fourth order ETH free cumulants defined in Eq.~\eqref{Eq:k}, respectively. This relation has previously been verified numerically in local chaotic systems \cite{PhysRevLett.134.140404} and has recently been shown to remain applicable in local integrable spin chains \cite{n7yh-828k}. Fig.~\ref{fig:all_model_OTOC} shows the ${\rm OTOC}(t)$ with the two cumulant contributions and their sum for the bilinear Majorana observable $\hat A=i\chi_1\chi_2$ in all three models (see Supplemental Material \cite{supp} for quartic Majorana observable and smaller system sizes). 

We find that Eq.~\eqref{eq:OTOC} holds approximately in all three models over the accessible time windows. At $t=0$, the positive second-order contribution $2[k_2^{\rm ETH}(0)]^2$ is compensated in part by the negative fourth order cumulant $k_4^{\rm ETH}(0)$, reproducing the initial ${\rm OTOC}(0)$. As time increases, $2[k_2^{\rm ETH}(t)]^2$ rapidly decays and becomes negligible, so that the post-transient ${\rm OTOC}(t)$ is predominantly governed by $k_4^{\rm ETH}(t)$. Here, late times refer to the post-transient regime of the finite-size calculations, where ${\rm OTOC}(t)$ and $k_4^{\rm ETH}(t)$ begin to follow similar dynamics. Remarkably, the three models exhibit distinctly different relaxation behavior. For chaotic $\rm SYK_4$ (Fig.~\ref{fig:all_model_OTOC}(a)), the ${\rm OTOC}(t)$ decays, exhibits a small negative transient, and approaches zero, together with $k_4^{\rm ETH}(t)$. In contrast, the integrable quadratic $\rm SYK_2$ model (Fig.~\ref{fig:all_model_OTOC}(b)) shows an initial rapid decrease of the ${\rm OTOC}(t)$ toward zero, followed by a recovery to a finite positive plateau. The $k_4^{\rm ETH}(t)$ changes sign and approaches a nearly stationary positive value $k_4^{\rm ETH}(t\gg2)\approx 0.70$, which determines the late-time ${\rm OTOC}(t)$. The integrable disorder-free quartic $\rm SYK_4$ (Fig.~\ref{fig:all_model_OTOC}(c)) exhibits a qualitatively different evolution: the ${\rm OTOC}(t)$ develops a pronounced negative minimum and subsequently undergoes a much slower recovery, eventually becoming positive at late times ${\rm OTOC}(t\gg20)\approx 0.15$. The $k_4^{\rm ETH}(t)$ follows the same nonmonotonic behavior and dominates the ${\rm OTOC}(t)$ after the  $2[k_2^{\rm ETH}(t)]^2$ become negligible. Notably, the positive late-time value is considerably smaller than in $\rm SYK_2$, and the relaxation occurs over a substantially longer timescale.

These observations extend the findings for local spin chains to all-to-all Majorana systems. In particular, the ${\rm OTOC}(t)$ decomposition (Eq.~\eqref{eq:OTOC}) remains approximately valid in both integrable models, although $k_4^{\rm ETH}(t)$ exhibits markedly different relaxation dynamics and late-time values, distinguishing not only chaotic from integrable behavior but also the two distinct integrable models. As emphasized in the full ETH framework, Eq.~\eqref{eq:OTOC} is expected when crossing contributions are suppressed and cactus contributions factorize. However, deviations from cactus factorization may have little visible effect on the ${\rm OTOC}(t)$ at late times, when $[k_2^{\rm ETH}(t)]^2$ is already small and $k_4^{\rm ETH}(t)$ dominates. Consequently, agreement with the ${\rm OTOC}(t)$ decomposition alone can not establish the validity of the full ETH conditions or distinguish chaotic from integrable dynamics. We therefore turn to the finite-size scaling of crossing contributions and deviations from cactus factorization. 

\begin{figure}[h!]
    \centering
    \includegraphics[width=\columnwidth]{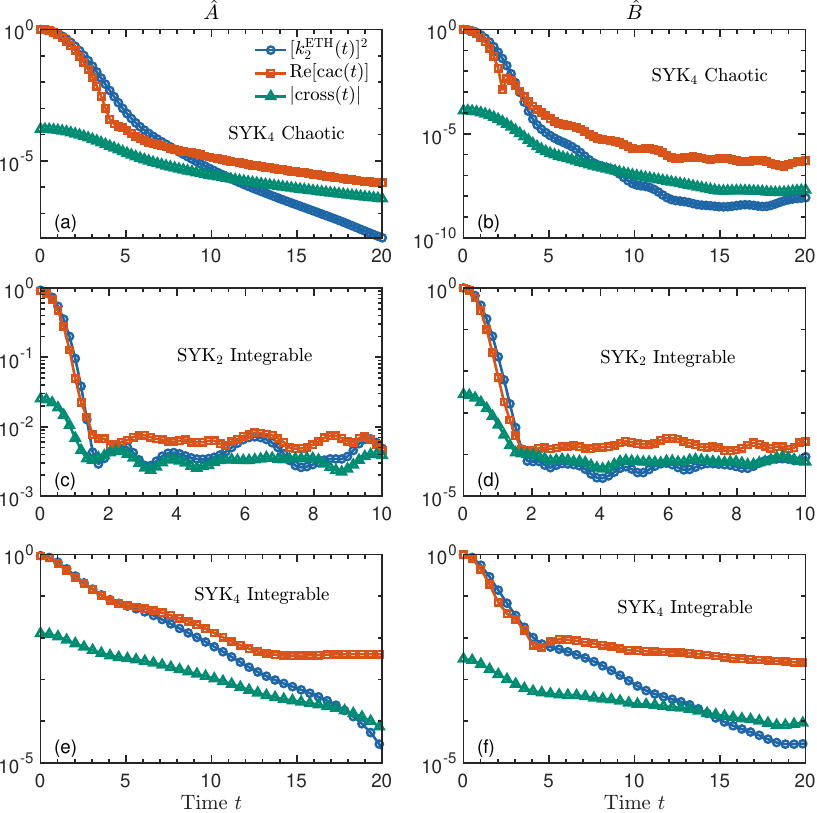}
    \caption{Time evolution of the cactus contribution $\mathrm{Re}[\mathrm{cac}(t)]$, compared with the factorized second-order ETH free cumulant $[k_2^{\rm ETH}(t)]^2$ and the crossing contribution $|\mathrm{cross}(t)|$, for (a,b) chaotic quartic $\rm SYK_{4}$, (c,d) integrable quadratic $\rm SYK_{2}$, and (e,f) integrable disorder-free quartic $\rm SYK_{4}$. The left and right columns correspond to the bilinear observable $\hat{A}=i\chi_1\chi_2$ and the quartic observable $\hat{B}=\chi_1\chi_2\chi_3\chi_4$, respectively. Results are shown for $N=30$, with 32 disorder realizations for the disordered models.}
    \label{fig:all_model_cactus_crossing}
\end{figure}

We next investigate crossing suppression and cactus factorization for the bilinear and quartic Majorana observables, $\hat A=i\chi_1\chi_2$ and $\hat B=\chi_1\chi_2\chi_3\chi_4$. Fig.~\ref{fig:all_model_cactus_crossing} compares $\mathrm{Re}[\mathrm{cac}(t)]$ with $[k_2^{\rm ETH}(t)]^2$, together with $|\mathrm{cross}(t)|$ for all the three models (see Supplemental Material \cite{supp} for smaller system sizes and cactus factorization in the frequency domain). Previous studies of local chaotic and integrable spin chains have found approximate cactus factorization at early times, followed by finite-size deviations whose magnitude and time dependence depend on integrability and the observable \cite{PhysRevLett.134.140404,n7yh-828k}.

For chaotic quartic $\rm SYK_{4}$ (Figs.~\ref{fig:all_model_cactus_crossing}(a,b)), the cactus contribution approximately follows the factorized result during the initial decay, up to $t\approx 2$-$3$, followed by visible deviations at later times. Both quantities decrease rapidly over several orders of magnitude, while the crossing contribution remains substantially smaller than the cactus contribution. This behavior is qualitatively consistent with the results for chaotic local spin chains, indicating that the suppression of crossing contributions and approximate early-time cactus factorization also occur in systems with all-to-all interactions.

The two integrable models exhibit distinctly different behavior. For the quadratic $\rm SYK_{2}$ (Figs.~\ref{fig:all_model_cactus_crossing}(c,d)), cactus factorization is approximately satisfied during the initial decay ($t\lesssim 1$), but noticeable deviations emerge around $t\approx 2$. Thereafter, both contributions exhibit persistent fluctuations around small finite values. The crossing contribution is considerably larger than in chaotic $\rm SYK_{4}$ and becomes comparable to the other contributions at later times. In contrast, integrable disorder-free quartic $\rm SYK_{4}$ (Figs.~\ref{fig:all_model_cactus_crossing}(e,f)) exhibit considerably slower relaxation. For the bilinear observable $\hat A$, the cactus and factorized contributions follow each other closely up to approximately $t\approx 5$-$6$, with deviations becoming increasingly apparent thereafter. For the quartic observable $\hat B$, the separation develops earlier, around $t\approx 3$-$4$. At later times ($t\gtrsim 10$), the cactus contributions decrease much more slowly and approach nearly stationary values, whereas the factorized contributions continue to decay. Thus, the apparent persistence of factorization in the disorder-free model is restricted to an intermediate time window and depends on the observable.    

These results extend the observations for local spin chains to all-to-all Majorana systems, while revealing distinct relaxation behavior within the integrable regime. In particular, unlike the rapid decay and subsequent fluctuations in quadratic $\rm SYK_{2}$, the interacting disorder-free $\rm SYK_{4}$ model exhibits prolonged relaxation and a stronger dependence on the observable. The comparison demonstrates that integrability alone does not determine the time dependence of higher-order matrix element correlations. The contrasting behavior of these two integrable Majorana models highlights the sensitivity of higher-order correlations to the microscopic interaction structure.

\begin{figure}[h!]
    \centering
    \includegraphics[width=\columnwidth]{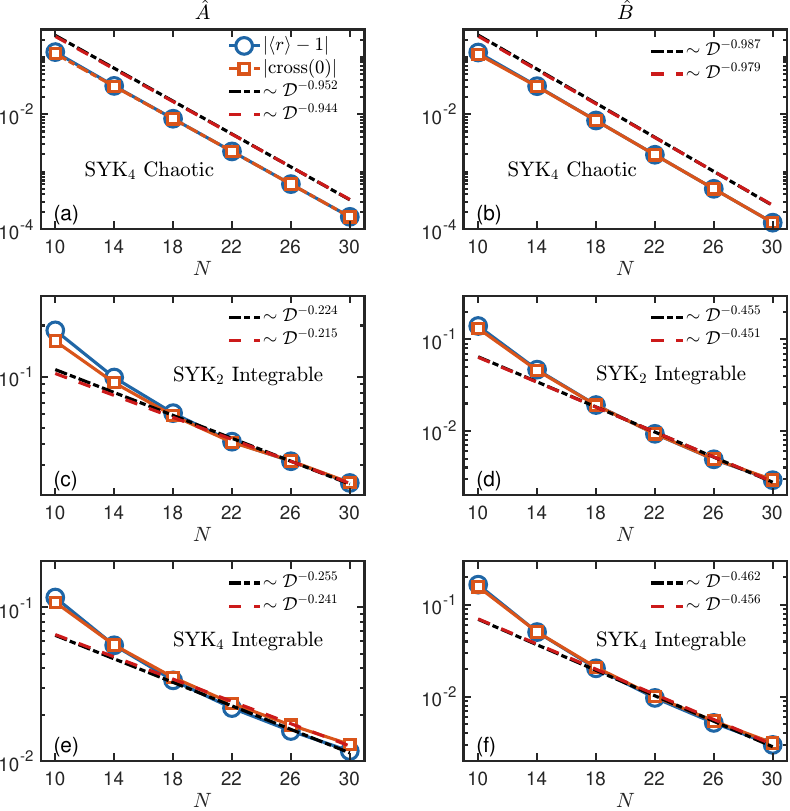}
    \caption{Finite-size scaling of the deviation from cactus factorization, $|\langle r\rangle-1|$, and the zero-time crossing contribution, $|\mathrm{cross}(0)|$, for (a,b) chaotic quartic $\rm SYK_{4}$, (c,d) integrable quadratic $\rm SYK_{2}$, and (e,f) integrable disorder-free quartic $\rm SYK_{4}$. The left and right columns correspond to the bilinear $\hat A=i\chi_1\chi_2$ and the quartic $\hat B=\chi_1\chi_2\chi_3\chi_4$ observable, respectively. Dashed lines indicate power-law fits proportional to $\sim\mathcal D^{-\gamma}$. The corresponding exponents are indicated in each panel; for the integrable models, the fits use the four largest system sizes.}
    \label{fig:all_model_scaling}
\end{figure}

To quantify finite-size deviations from cactus factorization, we consider the ratio \cite{PhysRevLett.134.140404}
\begin{equation}
\langle r\rangle=\frac{{\rm cac}(0,0)}{[k_2^{\rm ETH}(0)]^2}.
\label{eq}
\end{equation}
For the disordered models defined in Eqs.~\eqref{eq:SYK4C} and \eqref{eq:SYK2I}, the angular brackets denote an average over disorder realizations, whereas for the disorder-free model (Eq.~\eqref{eq:SYK4I}), the ratio is evaluated directly. Full ETH predicts $\langle r\rangle\rightarrow 1$ in the thermodynamic limit. We also examine the crossing contribution at $t=0$, where its magnitude is maximal. Fig.~\ref{fig:all_model_scaling} shows the system-size dependence of $|\langle r\rangle-1|$ and $|{\rm cross}(0)|$ for the bilinear $\hat A=i\chi_1\chi_2$ and quartic $\hat B=\chi_1\chi_2\chi_3\chi_4$ observable for all three models.

For chaotic quartic $\rm SYK_{4}$ (Figs.~\ref{fig:all_model_scaling}(a,b)), both quantities decrease rapidly with increasing Hilbert-space dimension $\mathcal D=2^{N/2-1}$. Power-law fits of the form $\sim\mathcal D^{-\gamma}$ yield exponents $\gamma\approx0.952$($0.944$) for the cactus factorization(crossing contribution) of $\hat A$ and $\gamma\approx0.987$($0.979$) for $\hat B$, indicating an approximately $\sim\mathcal D^{-1}$ suppression for both observables. Similar behavior has been reported for chaotic local spin chains \cite{PhysRevLett.134.140404}. Our results demonstrate that this characteristic suppression of higher-order matrix-element correlations also extends to chaotic systems with all-to-all interactions.  

In contrast, the two integrable models exhibit substantially slower suppression (Figs.~\ref{fig:all_model_scaling}(c)-(f))). For quadratic $\rm SYK_{2}$, fits to the four largest system sizes yield effective exponents $\gamma\approx0.224$($0.215$) for $\hat A$ and $\gamma\approx0.455$($0.451$) for $\hat B$. Remarkably, disorder-free quartic $\rm SYK_{4}$ model exhibits closely comparable values, $\gamma\approx0.255$($0.241$) and $\gamma\approx0.462$($0.456$). Previous study of local integrable spin chains have also revealed slower suppression of full ETH corrections, although the distinction between chaotic and integrable behavior can depend sensitively on the observable \cite{n7yh-828k}. In our all-to-all Majorana models, this distinction is clearly observed for both operators, with the quartic observable exhibiting approximately twice the effective exponent of the bilinear observable in the integrable cases. The near agreement between the two integrable models is particularly noteworthy: despite their different interaction structures and relaxation dynamics, they exhibit similar effective scaling. In both models, the crossing contributions and deviations from cactus factorization decrease substantially more slowly with increasing Hilbert-space dimension $\mathcal D$ than the approximately $\sim\mathcal D^{-1}$ behavior observed in chaotic SYK$_4$. Thus, the presence of four-body interactions alone does not guarantee the same finite-size suppression of higher-order matrix-element correlations as in the chaotic model. Interestingly, previous work has shown that the disorder-free integrable SYK$_4$ model can exhibit short-time exponential-like growth of OTOCs at finite temperature, despite its integrability \cite{PhysRevResearch.7.013092}. Our results provide a complementary perspective: while the approximate OTOC cumulant decomposition is observed in both chaotic and integrable models (Fig.~\ref{fig:all_model_OTOC}), the finite-size scaling of crossing contributions and deviations from cactus factorization (Fig.~\ref{fig:all_model_scaling}) reveals a clear distinction between them. We emphasize that the effective exponents obtained for the integrable models characterize the accessible finite-size range, and their asymptotic behavior remains to be established.

\emph{Outlook and discussion.} We have investigated full ETH in chaotic and integrable all-to-all Majorana systems. For chaotic quartic $\rm SYK_{4}$, we find that the approximate ${\rm OTOC}(t)$ decomposition is well described by the second and fourth-order ETH free cumulants, while crossing contributions and deviations from cactus factorization are suppressed approximately as $\sim\mathcal D^{-1}$. This is consistent with the full ETH behavior previously observed in chaotic local systems and demonstrates that its characteristic finite-size suppression also emerges in all-to-all Majorana models. 

Interestingly, the approximate ${\rm OTOC}(t)$ decomposition remains applicable in both integrable quadratic $\rm SYK_{2}$ and integrable disorder-free quartic $\rm SYK_{4}$ models, indicating that this relation alone does not distinguish chaotic from integrable dynamics. Nevertheless, the fourth-order ETH free cumulant $k_4^{\rm ETH}(t)$ captured their markedly different relaxation behavior. In chaotic $\rm SYK_{4}$, it rapidly approaches zero, whereas in integrable $\rm SYK_{2}$, it approaches a finite stationary value. In contrast, the integrable disorder-free $\rm SYK_{4}$ exhibits much slower, non-monotonic relaxation toward a considerably smaller positive value. Thus, although $k_4^{\rm ETH}(t)$ governs the post-transient ${\rm OTOC}(t)$ in all three models, its dynamics distinguishes not only chaotic from integrable behavior but also the relaxation of noninteracting and interacting integrable systems.

A clear distinction in the full ETH conditions emerges from their finite-size corrections. Unlike the approximately $\sim\mathcal D^{-1}$ suppression in chaotic $\rm SYK_{4}$, both integrable models exhibit substantially slower, observable dependent effective scaling. Notably, the quadratic $\rm SYK_{2}$ and interacting disorder-free $\rm SYK_{4}$ models display closely comparable exponents despite their different interaction structures and relaxation dynamics. This indicates that the presence of four-body interactions alone does not ensure the $\sim\mathcal D^{-1}$ suppression of crossing contributions and deviations from cactus factorization observed in chaotic SYK$_4$. Our findings therefore establish a qualitative distinction between chaotic and integrable full-ETH behavior beyond spatially local systems. An important open question is whether the effective scaling exponents observed in the integrable models persist at larger system sizes or crossover to algebraic scaling in $N$. Developing an analytical understanding of these finite-size corrections and the distinct relaxation dynamics of $k_4^{\rm ETH}(t)$ would help clarify the behavior of higher-order correlations in integrable systems. Another promising direction is to investigate whether the scaling of crossing contributions and deviations from cactus factorization can serve as a diagnostic of ergodicity-breaking transitions, particularly in models such as the Rosenzweig-Porter model \cite{Kravtsov2015RP,Buijsman2024RP,Jahnke2025RP,PhysRevE.111.024201}, which exhibits ergodic, nonergodic extended, and localized regimes. Extending these investigations to higher-order ETH free cumulants and other classes of all-to-all many-body Hamiltonians \cite{Hanada2024Spins,Saha2026Scrambling,9jbx-9kln} could further clarify the conditions under which the full-ETH structure emerges beyond conventional thermalization.

\emph{Acknowledgments.---} A.~K.~S. acknowledges support from JSPS KAKENHI Grant Number JP25K00925. The author acknowledges the computational facilities of the Yukawa Institute for Theoretical Physics. The author also thanks Tanay Pathak for valuable discussions and critical reading of the manuscript.

\bibliography{references}

@misc{supp,
  note = "See Supplemental Material at
    URL-will-be-inserted-by-publisher for details on numerics and additional supporting results."
}

@article{n7yh-828k,
  title = {Full eigenstate thermalization in integrable spin systems},
  author = {Pathak, Tanay},
  journal = {Phys. Rev. E},
  volume = {114},
  issue = {4},
  pages = {044124},
  numpages = {12},
  year = {2026},
  month = {Oct},
  publisher = {American Physical Society},
  doi = {10.1103/n7yh-828k},
  url = {https://link.aps.org/doi/10.1103/n7yh-828k}
}

@article{PhysRevB.100.115122,
  title = {Eigenstate thermalization scaling in Majorana clusters: From chaotic to integrable Sachdev-Ye-Kitaev models},
  author = {Haque, Masudul and McClarty, P. A.},
  journal = {Phys. Rev. B},
  volume = {100},
  issue = {11},
  pages = {115122},
  numpages = {5},
  year = {2019},
  month = {Sep},
  publisher = {American Physical Society},
  doi = {10.1103/PhysRevB.100.115122},
  url = {https://link.aps.org/doi/10.1103/PhysRevB.100.115122}
}

@article{PhysRevE.60.3949,
  title = {Ergodic properties of a generic nonintegrable quantum many-body system in the thermodynamic limit},
  author = {Prosen, Toma{\ifmmode \check{z}\else \v{z}\fi{}}},
  journal = {Phys. Rev. E},
  volume = {60},
  issue = {4},
  pages = {3949--3968},
  numpages = {0},
  year = {1999},
  month = {Oct},
  publisher = {American Physical Society},
  doi = {10.1103/PhysRevE.60.3949},
  url = {https://link.aps.org/doi/10.1103/PhysRevE.60.3949}
}

@article{Rigol_2008,
   title={Thermalization and its mechanism for generic isolated quantum systems},
   volume={452},
   ISSN={1476-4687},
   url={http://dx.doi.org/10.1038/nature06838},
   DOI={10.1038/nature06838},
   number={7189},
   journal={Nature},
   publisher={Springer Science and Business Media LLC},
   author={Rigol, Marcos and Dunjko, Vanja and Olshanii, Maxim},
   year={2008},
   month=Apr, pages={854–858} }

@article{PhysRevLett.105.250401,
  title = {Effect of Rare Fluctuations on the Thermalization of Isolated Quantum Systems},
  author = {Biroli, Giulio and Kollath, Corinna and L{\"a}uchli, Andreas M.},
  journal = {Phys. Rev. Lett.},
  volume = {105},
  issue = {25},
  pages = {250401},
  numpages = {4},
  year = {2010},
  month = {Dec},
  publisher = {American Physical Society},
  doi = {10.1103/PhysRevLett.105.250401},
  url = {https://link.aps.org/doi/10.1103/PhysRevLett.105.250401}
}

@article{RevModPhys.83.863,
  title = {Colloquium: Nonequilibrium dynamics of closed interacting quantum systems},
  author = {Polkovnikov, Anatoli and Sengupta, Krishnendu and Silva, Alessandro and Vengalattore, Mukund},
  journal = {Rev. Mod. Phys.},
  volume = {83},
  issue = {3},
  pages = {863--883},
  numpages = {0},
  year = {2011},
  month = {Aug},
  publisher = {American Physical Society},
  doi = {10.1103/RevModPhys.83.863},
  url = {https://link.aps.org/doi/10.1103/RevModPhys.83.863}
}

@article{PhysRevE.87.012125,
  title = {Finite-size scaling analysis of the eigenstate thermalization hypothesis in a one-dimensional interacting Bose gas},
  author = {Ikeda, Tatsuhiko N. and Watanabe, Yu and Ueda, Masahito},
  journal = {Phys. Rev. E},
  volume = {87},
  issue = {1},
  pages = {012125},
  numpages = {5},
  year = {2013},
  month = {Jan},
  publisher = {American Physical Society},
  doi = {10.1103/PhysRevE.87.012125},
  url = {https://link.aps.org/doi/10.1103/PhysRevE.87.012125}
}

@article{D_Alessio_2016,
   title={From quantum chaos and eigenstate thermalization to statistical mechanics and thermodynamics},
   volume={65},
   ISSN={1460-6976},
   url={http://dx.doi.org/10.1080/00018732.2016.1198134},
   DOI={10.1080/00018732.2016.1198134},
   number={3},
   journal={Advances in Physics},
   publisher={Informa UK Limited},
   author={D’Alessio, Luca and Kafri, Yariv and Polkovnikov, Anatoli and Rigol, Marcos},
   year={2016},
   month=May, pages={239–362} }

@article{PhysRevLett.117.170404,
  title = {Anomalous Thermalization in Ergodic Systems},
  author = {Luitz, David J. and Bar Lev, Yevgeny},
  journal = {Phys. Rev. Lett.},
  volume = {117},
  issue = {17},
  pages = {170404},
  numpages = {5},
  year = {2016},
  month = {Oct},
  publisher = {American Physical Society},
  doi = {10.1103/PhysRevLett.117.170404},
  url = {https://link.aps.org/doi/10.1103/PhysRevLett.117.170404}
}

@article{PhysRevB.93.134201,
  title = {Long tail distributions near the many-body localization transition},
  author = {Luitz, David J.},
  journal = {Phys. Rev. B},
  volume = {93},
  issue = {13},
  pages = {134201},
  numpages = {10},
  year = {2016},
  month = {Apr},
  publisher = {American Physical Society},
  doi = {10.1103/PhysRevB.93.134201},
  url = {https://link.aps.org/doi/10.1103/PhysRevB.93.134201}
}

@article{PhysRevLett.129.030602,
  title = {Eigenstate Thermalization in Long-Range Interacting Systems},
  author = {Sugimoto, Shoki and Hamazaki, Ryusuke and Ueda, Masahito},
  journal = {Phys. Rev. Lett.},
  volume = {129},
  issue = {3},
  pages = {030602},
  numpages = {8},
  year = {2022},
  month = {Jul},
  publisher = {American Physical Society},
  doi = {10.1103/PhysRevLett.129.030602},
  url = {https://link.aps.org/doi/10.1103/PhysRevLett.129.030602}
}

@article{PhysRevE.99.042139,
  title = {Eigenstate thermalization hypothesis and out of time order correlators},
  author = {Foini, Laura and Kurchan, Jorge},
  journal = {Phys. Rev. E},
  volume = {99},
  issue = {4},
  pages = {042139},
  numpages = {12},
  year = {2019},
  month = {Apr},
  publisher = {American Physical Society},
  doi = {10.1103/PhysRevE.99.042139},
  url = {https://link.aps.org/doi/10.1103/PhysRevE.99.042139}
}

@article{PhysRevLett.129.170603,
  title = {Eigenstate Thermalization Hypothesis and Free Probability},
  author = {Pappalardi, Silvia and Foini, Laura and Kurchan, Jorge},
  journal = {Phys. Rev. Lett.},
  volume = {129},
  issue = {17},
  pages = {170603},
  numpages = {6},
  year = {2022},
  month = {Oct},
  publisher = {American Physical Society},
  doi = {10.1103/PhysRevLett.129.170603},
  url = {https://link.aps.org/doi/10.1103/PhysRevLett.129.170603}
}

@article{PhysRevLett.123.260601,
  title = {Eigenstate Thermalization and Rotational Invariance in Ergodic Quantum Systems},
  author = {Foini, Laura and Kurchan, Jorge},
  journal = {Phys. Rev. Lett.},
  volume = {123},
  issue = {26},
  pages = {260601},
  numpages = {4},
  year = {2019},
  month = {Dec},
  publisher = {American Physical Society},
  doi = {10.1103/PhysRevLett.123.260601},
  url = {https://link.aps.org/doi/10.1103/PhysRevLett.123.260601}
}

@article{PhysRevLett.122.220601,
  title = {Eigenstate Correlations, Thermalization, and the Butterfly Effect},
  author = {Chan, Amos and De Luca, Andrea and Chalker, J. T.},
  journal = {Phys. Rev. Lett.},
  volume = {122},
  issue = {22},
  pages = {220601},
  numpages = {5},
  year = {2019},
  month = {Jun},
  publisher = {American Physical Society},
  doi = {10.1103/PhysRevLett.122.220601},
  url = {https://link.aps.org/doi/10.1103/PhysRevLett.122.220601}
}

@article{PhysRevLett.123.230606,
  title = {Bounds on Chaos from the Eigenstate Thermalization Hypothesis},
  author = {Murthy, Chaitanya and Srednicki, Mark},
  journal = {Phys. Rev. Lett.},
  volume = {123},
  issue = {23},
  pages = {230606},
  numpages = {5},
  year = {2019},
  month = {Dec},
  publisher = {American Physical Society},
  doi = {10.1103/PhysRevLett.123.230606},
  url = {https://link.aps.org/doi/10.1103/PhysRevLett.123.230606}
}

@article{PhysRevE.104.034120,
  title = {Out-of-time-order correlations and the fine structure of eigenstate thermalization},
  author = {Brenes, Marlon and Pappalardi, Silvia and Mitchison, Mark T. and Goold, John and Silva, Alessandro},
  journal = {Phys. Rev. E},
  volume = {104},
  issue = {3},
  pages = {034120},
  numpages = {13},
  year = {2021},
  month = {Sep},
  publisher = {American Physical Society},
  doi = {10.1103/PhysRevE.104.034120},
  url = {https://link.aps.org/doi/10.1103/PhysRevE.104.034120}
}

@article{PhysRevLett.134.140404,
  title = {Full Eigenstate Thermalization via Free Cumulants in Quantum Lattice Systems},
  author = {Pappalardi, Silvia and Fritzsch, Felix and Prosen, Toma{\ifmmode \check{z}\else \v{z}\fi{}}},
  journal = {Phys. Rev. Lett.},
  volume = {134},
  issue = {14},
  pages = {140404},
  numpages = {7},
  year = {2025},
  month = {Apr},
  publisher = {American Physical Society},
  doi = {10.1103/PhysRevLett.134.140404},
  url = {https://link.aps.org/doi/10.1103/PhysRevLett.134.140404}
}

@article{PhysRevResearch.7.013092,
  title = {Disorder-free Sachdev-Ye-Kitaev models: Integrability and a precursor of chaos},
  author = {Ozaki, Soshun and Katsura, Hosho},
  journal = {Phys. Rev. Res.},
  volume = {7},
  issue = {1},
  pages = {013092},
  numpages = {16},
  year = {2025},
  month = {Jan},
  publisher = {American Physical Society},
  doi = {10.1103/PhysRevResearch.7.013092},
  url = {https://link.aps.org/doi/10.1103/PhysRevResearch.7.013092}
}

@article{PhysRevLett.70.3339,
  title = {{Gapless spin-fluid ground state in a random quantum Heisenberg magnet}},
  author = {Sachdev, Subir and Ye, Jinwu},
  journal = {Phys. Rev. Lett.},
  volume = {70},
  issue = {21},
  pages = {3339--3342},
  numpages = {0},
  year = {1993},
  month = {May},
  publisher = {American Physical Society},
  doi = {10.1103/PhysRevLett.70.3339},
  url = {https://link.aps.org/doi/10.1103/PhysRevLett.70.3339}
}

@misc{kitaev1998,
author       = {Alexei Kitaev},
title        = {{Hidden correlations in the Hawking radiation and thermal noise, talk at KITP, 2015}},
year         = {2015},
url         ={https://online.kitp.ucsb.edu/online/joint98/kitaev/}
}

@misc{kitaev2015,
author       = {Alexei Kitaev},
title        = {A simple model of quantum holography},
year         = {2015},
howpublished = {Talks at KITP},
note = {Available online: \url{http://online.kitp.ucsb.edu/online/entangled15/kitaev/}\,and \url{http://online.kitp.ucsb.edu/online/entangled15/kitaev2/}}
}

@article{PhysRevLett.120.241603,
  title = {Chaotic-Integrable Transition in the Sachdev-Ye-Kitaev Model},
  author = {Garc{\'{\i}}a-Garc{\'{\i}}a, Antonio M. and Loureiro, Bruno and Romero-Berm{\'u}dez, Aurelio and Tezuka, Masaki},
  journal = {Phys. Rev. Lett.},
  volume = {120},
  issue = {24},
  pages = {241603},
  numpages = {6},
  year = {2018},
  month = {Jun},
  publisher = {American Physical Society},
  doi = {10.1103/PhysRevLett.120.241603},
  url = {https://link.aps.org/doi/10.1103/PhysRevLett.120.241603}
}

@article{PhysRevB.95.115150,
  title = {{S}achdev-{Y}e-{K}itaev model and thermalization on the boundary of many-body localized fermionic symmetry-protected topological states},
  author = {You, Yi-Zhuang and Ludwig, Andreas W. W. and Xu, Cenke},
  journal = {Phys. Rev. B},
  volume = {95},
  issue = {11},
  pages = {115150},
  numpages = {12},
  year = {2017},
  month = {Mar},
  publisher = {American Physical Society},
  doi = {10.1103/PhysRevB.95.115150},
  url = {https://link.aps.org/doi/10.1103/PhysRevB.95.115150}
}

@article{PhysRevD.94.126010,
  title = {Spectral and thermodynamic properties of the Sachdev-Ye-Kitaev model},
  author = {Garc{\'{\i}}a-Garc{\'{\i}}a, Antonio M. and Verbaarschot, Jacobus J. M.},
  journal = {Phys. Rev. D},
  volume = {94},
  issue = {12},
  pages = {126010},
  numpages = {13},
  year = {2016},
  month = {Dec},
  publisher = {American Physical Society},
  doi = {10.1103/PhysRevD.94.126010},
  url = {https://link.aps.org/doi/10.1103/PhysRevD.94.126010}
}

@article{PhysRevD.94.106002,
  title = {Remarks on the Sachdev-Ye-Kitaev model},
  author = {Maldacena, Juan and Stanford, Douglas},
  journal = {Phys. Rev. D},
  volume = {94},
  issue = {10},
  pages = {106002},
  numpages = {43},
  year = {2016},
  month = {Nov},
  publisher = {American Physical Society},
  doi = {10.1103/PhysRevD.94.106002},
  url = {https://link.aps.org/doi/10.1103/PhysRevD.94.106002}
}

@article{PhysRevA.43.2046,
  title = {Quantum statistical mechanics in a closed system},
  author = {Deutsch, J. M.},
  journal = {Phys. Rev. A},
  volume = {43},
  issue = {4},
  pages = {2046--2049},
  numpages = {0},
  year = {1991},
  month = {Feb},
  publisher = {American Physical Society},
  doi = {10.1103/PhysRevA.43.2046},
  url = {https://link.aps.org/doi/10.1103/PhysRevA.43.2046}
}

@article{PhysRevE.50.888,
  title = {Chaos and quantum thermalization},
  author = {Srednicki, Mark},
  journal = {Phys. Rev. E},
  volume = {50},
  issue = {2},
  pages = {888--901},
  numpages = {0},
  year = {1994},
  month = {Aug},
  publisher = {American Physical Society},
  doi = {10.1103/PhysRevE.50.888},
  url = {https://link.aps.org/doi/10.1103/PhysRevE.50.888}
}

@misc{deutsch2025closedquantumgivingergodicity,
      title={A closed quantum system giving ergodicity}, 
      author={J. M. Deutsch},
      year={2025},
      eprint={2502.15947},
      archivePrefix={arXiv},
      primaryClass={quant-ph},
      url={https://arxiv.org/abs/2502.15947}, 
}

@article{d1cq-gj9p,
  title = {Refinements of the Eigenstate Thermalization Hypothesis under Local Rotational Invariance via Free Probability},
  author = {Vallini, Elisa and Foini, Laura and Pappalardi, Silvia},
  journal = {PRX Quantum},
  volume = {7},
  issue = {4},
  pages = {043007},
  numpages = {37},
  year = {2026},
  month = {Oct},
  publisher = {American Physical Society},
  doi = {10.1103/d1cq-gj9p},
  url = {https://link.aps.org/doi/10.1103/d1cq-gj9p}
}

@article{Vallini_2026,
   title={Long-time Freeness in the Kicked Top},
   volume={10},
   ISSN={2521-327X},
   url={http://dx.doi.org/10.22331/q-2026-06-08-2129},
   DOI={10.22331/q-2026-06-08-2129},
   journal={Quantum},
   publisher={Verein zur Forderung des Open Access Publizierens in den Quantenwissenschaften},
   author={Vallini, Elisa and Pappalardi, Silvia},
   year={2026},
   month=June, pages={2129} }

@misc{zhang2026finitesizescalingeigenstatethermalization,
      title={Finite-Size Scaling of the Full Eigenstate Thermalization in Quantum Spin Chains}, 
      author={Yuke Zhang and Pengfei Zhang},
      year={2026},
      eprint={2602.01809},
      archivePrefix={arXiv},
      primaryClass={quant-ph},
      url={https://arxiv.org/abs/2602.01809}, 
}

@article{PhysRevX.5.041025,
  title = {Bekenstein-Hawking Entropy and Strange Metals},
  author = {Sachdev, Subir},
  journal = {Phys. Rev. X},
  volume = {5},
  issue = {4},
  pages = {041025},
  numpages = {13},
  year = {2015},
  month = {Nov},
  publisher = {American Physical Society},
  doi = {10.1103/PhysRevX.5.041025},
  url = {https://link.aps.org/doi/10.1103/PhysRevX.5.041025}
}

@article{PhysRevB.105.075117,
  title = {Thermalization of many many-body interacting Sachdev-Ye-Kitaev models},
  author = {Louw, Jan C. and Kehrein, Stefan},
  journal = {Phys. Rev. B},
  volume = {105},
  issue = {7},
  pages = {075117},
  numpages = {8},
  year = {2022},
  month = {Feb},
  publisher = {American Physical Society},
  doi = {10.1103/PhysRevB.105.075117},
  url = {https://link.aps.org/doi/10.1103/PhysRevB.105.075117}
}

@article{PhysRevB.105.235131,
  title = {Near-equilibrium approach to transport in complex Sachdev-Ye-Kitaev models},
  author = {Zanoci, Cristian and Swingle, Brian},
  journal = {Phys. Rev. B},
  volume = {105},
  issue = {23},
  pages = {235131},
  numpages = {21},
  year = {2022},
  month = {Jun},
  publisher = {American Physical Society},
  doi = {10.1103/PhysRevB.105.235131},
  url = {https://link.aps.org/doi/10.1103/PhysRevB.105.235131}
}

@article{PhysRevB.94.035135,
  title = {Numerical study of fermion and boson models with infinite-range random interactions},
  author = {Fu, Wenbo and Sachdev, Subir},
  journal = {Phys. Rev. B},
  volume = {94},
  issue = {3},
  pages = {035135},
  numpages = {9},
  year = {2016},
  month = {Jul},
  publisher = {American Physical Society},
  doi = {10.1103/PhysRevB.94.035135},
  url = {https://link.aps.org/doi/10.1103/PhysRevB.94.035135}
}

@article{PhysRevB.100.155128,
  title = {Chaos in a classical limit of the Sachdev-Ye-Kitaev model},
  author = {Scaffidi, Thomas and Altman, Ehud},
  journal = {Phys. Rev. B},
  volume = {100},
  issue = {15},
  pages = {155128},
  numpages = {5},
  year = {2019},
  month = {Oct},
  publisher = {American Physical Society},
  doi = {10.1103/PhysRevB.100.155128},
  url = {https://link.aps.org/doi/10.1103/PhysRevB.100.155128}
}

@article{Gross_2017,
  title   = {A Generalization of Sachdev-Ye-Kitaev},
  author  = {Gross, David J. and Rosenhaus, Vladimir},
  journal = {Journal of High Energy Physics},
  volume  = {2017},
  number  = {2},
  pages   = {093},
  year    = {2017},
  doi     = {10.1007/JHEP02(2017)093},
  url     = {https://doi.org/10.1007/JHEP02(2017)093}
}

@article{Li_2017,
  title   = {Supersymmetric SYK Model and Random Matrix Theory},
  author  = {Li, Tianlin and Liu, Junyu and Xin, Yuan and Zhou, Yehao},
  journal = {Journal of High Energy Physics},
  volume  = {2017},
  number  = {6},
  pages   = {111},
  year    = {2017},
  doi     = {10.1007/JHEP06(2017)111},
  url     = {https://doi.org/10.1007/JHEP06(2017)111}
}

@article{Fu_2017,
  title   = {Supersymmetric Sachdev-Ye-Kitaev Models},
  author  = {Fu, Wenbo and Gaiotto, Davide and Maldacena, Juan and Sachdev, Subir},
  journal = {Phys. Rev. D},
  volume  = {95},
  number  = {2},
  pages   = {026009},
  year    = {2017},
  doi     = {10.1103/PhysRevD.95.026009},
  url     = {https://doi.org/10.1103/PhysRevD.95.026009}
}

@article{PhysRevX.12.021040,
  title = {Symmetry Classification and Universality in Non-Hermitian Many-Body Quantum Chaos by the Sachdev-Ye-Kitaev Model},
  author = {Garc{\'{\i}}a-Garc{\'{\i}}a, Antonio M. and S{\'a}, Lucas and Verbaarschot, Jacobus J. M.},
  journal = {Phys. Rev. X},
  volume = {12},
  issue = {2},
  pages = {021040},
  numpages = {22},
  year = {2022},
  month = {May},
  publisher = {American Physical Society},
  doi = {10.1103/PhysRevX.12.021040},
  url = {https://link.aps.org/doi/10.1103/PhysRevX.12.021040}
}

@article{Gu_2017,
  title   = {Local Criticality, Diffusion and Chaos in Generalized Sachdev-Ye-Kitaev Models},
  author  = {Gu, Yingfei and Qi, Xiao-Liang and Stanford, Douglas},
  journal = {Journal of High Energy Physics},
  volume  = {2017},
  number  = {5},
  pages   = {125},
  year    = {2017},
  doi     = {10.1007/JHEP05(2017)125},
  url     = {https://doi.org/10.1007/JHEP05(2017)125}
}

@article{Berkooz_2018,
  title   = {Chord Diagrams, Exact Correlators in Spin Glasses and Black Hole Bulk Reconstruction},
  author  = {Berkooz, Micha and Narayan, Prithvi and Sim{\'o}n, Joan},
  journal = {Journal of High Energy Physics},
  volume  = {2018},
  number  = {8},
  pages   = {192},
  year    = {2018},
  doi     = {10.1007/JHEP08(2018)192},
  url     = {https://doi.org/10.1007/JHEP08(2018)192}
}

@article{PhysRevLett.105.151602,
  title = {Holographic Metals and the Fractionalized Fermi Liquid},
  author = {Sachdev, Subir},
  journal = {Phys. Rev. Lett.},
  volume = {105},
  issue = {15},
  pages = {151602},
  numpages = {4},
  year = {2010},
  month = {Oct},
  publisher = {American Physical Society},
  doi = {10.1103/PhysRevLett.105.151602},
  url = {https://link.aps.org/doi/10.1103/PhysRevLett.105.151602}
}

@article{Cotler_2017,
  title   = {Black Holes and Random Matrices},
  author  = {Cotler, Jordan S. and Gur-Ari, Guy and Hanada, Masanori and Polchinski, Joseph and Saad, Phil and Shenker, Stephen H. and Stanford, Douglas and Streicher, Alexandre and Tezuka, Masaki},
  journal = {Journal of High Energy Physics},
  volume  = {2017},
  number  = {5},
  pages   = {118},
  year    = {2017},
  doi     = {10.1007/JHEP05(2017)118},
  url     = {https://doi.org/10.1007/JHEP05(2017)118}
}

@article{RevModPhys.94.035004,
  title = {Sachdev-Ye-Kitaev models and beyond: Window into non-Fermi liquids},
  author = {Chowdhury, Debanjan and Georges, Antoine and Parcollet, Olivier and Sachdev, Subir},
  journal = {Rev. Mod. Phys.},
  volume = {94},
  issue = {3},
  pages = {035004},
  numpages = {78},
  year = {2022},
  month = {Sep},
  publisher = {American Physical Society},
  doi = {10.1103/RevModPhys.94.035004},
  url = {https://link.aps.org/doi/10.1103/RevModPhys.94.035004}
}

@misc{faulkner2022snowmasswhitepaperquantum,
  title         = {Snowmass White Paper: Quantum Information in Quantum Field Theory and Quantum Gravity},
  author        = {Faulkner, Thomas and Hartman, Thomas and Headrick, Matthew and Rangamani, Mukund and Swingle, Brian},
  year          = {2022},
  eprint        = {2203.07117},
  archivePrefix = {arXiv},
  primaryClass  = {hep-th},
  url           = {https://arxiv.org/abs/2203.07117}
}

@article{PhysRevE.102.042127,
  title = {Eigenstate thermalization hypothesis beyond standard indicators: Emergence of random-matrix behavior at small frequencies},
  author = {Richter, Jonas and Dymarsky, Anatoly and Steinigeweg, Robin and Gemmer, Jochen},
  journal = {Phys. Rev. E},
  volume = {102},
  issue = {4},
  pages = {042127},
  numpages = {12},
  year = {2020},
  month = {Oct},
  publisher = {American Physical Society},
  doi = {10.1103/PhysRevE.102.042127},
  url = {https://link.aps.org/doi/10.1103/PhysRevE.102.042127}
}

@article{Hashimoto_2017,
  title   = {Out-of-time-order correlators in quantum mechanics},
  author  = {Hashimoto, Koji and Murata, Keiju and Yoshii, Ryosuke},
  journal = {Journal of High Energy Physics},
  volume  = {2017},
  number  = {10},
  pages   = {138},
  year    = {2017},
  doi     = {10.1007/JHEP10(2017)138},
  url     = {https://doi.org/10.1007/JHEP10(2017)138}
}

@article{Roberts_2017,
  title   = {Chaos and Complexity by Design},
  author  = {Roberts, Daniel A. and Yoshida, Beni},
  journal = {Journal of High Energy Physics},
  volume  = {2017},
  number  = {4},
  pages   = {121},
  year    = {2017},
  doi     = {10.1007/JHEP04(2017)121},
  url     = {https://doi.org/10.1007/JHEP04(2017)121}
}

@article{Hosur_2016,
  title   = {Chaos in Quantum Channels},
  author  = {Hosur, Pavan and Qi, Xiao-Liang and Roberts, Daniel A. and Yoshida, Beni},
  journal = {Journal of High Energy Physics},
  volume  = {2016},
  number  = {2},
  pages   = {004},
  year    = {2016},
  doi     = {10.1007/JHEP02(2016)004},
  url     = {https://doi.org/10.1007/JHEP02(2016)004}
}

@article{Maldacena_2016,
  title   = {A Bound on Chaos},
  author  = {Maldacena, Juan and Shenker, Stephen H. and Stanford, Douglas},
  journal = {Journal of High Energy Physics},
  volume  = {2016},
  number  = {8},
  pages   = {106},
  year    = {2016},
  doi     = {10.1007/JHEP08(2016)106},
  url     = {https://doi.org/10.1007/JHEP08(2016)106}
}

@article{PhysRevB.111.054303,
  title = {Microcanonical free cumulants in lattice systems},
  author = {Fritzsch, Felix and Prosen, Toma{\ifmmode \check{z}\else \v{z}\fi{}} and Pappalardi, Silvia},
  journal = {Phys. Rev. B},
  volume = {111},
  issue = {5},
  pages = {054303},
  numpages = {16},
  year = {2025},
  month = {Feb},
  publisher = {American Physical Society},
  doi = {10.1103/PhysRevB.111.054303},
  url = {https://link.aps.org/doi/10.1103/PhysRevB.111.054303}
}

@article{6n9v-b3b7,
  title = {Free Cumulants in Full Eigenstate Thermalization from Boundary Scrambling},
  author = {Fritzsch, Felix and Alves, Gabriel O. and Rampp, Michael A. and Claeys, Pieter W.},
  journal = {PRX Quantum},
  volume = {7},
  issue = {4},
  pages = {043003},
  numpages = {26},
  year = {2026},
  month = {Oct},
  publisher = {American Physical Society},
  doi = {10.1103/6n9v-b3b7},
  url = {https://link.aps.org/doi/10.1103/6n9v-b3b7}
}

@article{PhysRevLett.125.250602,
  title = {Exponential Ramp in the Quadratic Sachdev-Ye-Kitaev Model},
  author = {Winer, Michael and Jian, Shao-Kai and Swingle, Brian},
  journal = {Phys. Rev. Lett.},
  volume = {125},
  issue = {25},
  pages = {250602},
  numpages = {6},
  year = {2020},
  month = {Dec},
  publisher = {American Physical Society},
  doi = {10.1103/PhysRevLett.125.250602},
  url = {https://link.aps.org/doi/10.1103/PhysRevLett.125.250602}
}

@article{Kravtsov2015RP,
  title   = {A Random Matrix Model with Localization and Ergodic Transitions},
  author  = {Kravtsov, V. E. and Khaymovich, I. M. and Cuevas, E. and Amini, M.},
  journal = {New Journal of Physics},
  volume  = {17},
  pages   = {122002},
  year    = {2015},
  doi     = {10.1088/1367-2630/17/12/122002},
  url     = {https://doi.org/10.1088/1367-2630/17/12/122002}
}

@article{Buijsman2024RP,
  title   = {Long-Range Spectral Statistics of the Rosenzweig-Porter Model},
  author  = {Buijsman, Wouter},
  journal = {Phys. Rev. B},
  volume  = {109},
  pages   = {024205},
  year    = {2024},
  doi     = {10.1103/PhysRevB.109.024205},
  url     = {https://doi.org/10.1103/PhysRevB.109.024205}
}

@article{Jahnke2025RP,
  title   = {Free Probability Approach to Spectral and Operator Statistics in Rosenzweig-Porter Random Matrix Ensembles},
  author  = {Jahnke, Viktor and Nandy, Pratik and Pal, Kuntal and Camargo, Hugo A. and Kim, Keun-Young},
  journal = {Journal of High Energy Physics},
  volume  = {2025},
  number  = {12},
  pages   = {002},
  year    = {2025},
  doi     = {10.1007/JHEP12(2025)002},
  url     = {https://doi.org/10.1007/JHEP12(2025)002}
}

@article{Hanada2024Spins,
  title   = {A Model of Randomly-Coupled Pauli Spins},
  author  = {Hanada, Masanori and Jevicki, Antal and Liu, Xianlong and Rinaldi, Enrico and Tezuka, Masaki},
  journal = {Journal of High Energy Physics},
  volume  = {2024},
  number  = {5},
  pages   = {280},
  year    = {2024},
  doi     = {10.1007/JHEP05(2024)280},
  url     = {https://doi.org/10.1007/JHEP05(2024)280}
}

@misc{Saha2026Scrambling,
  title         = {Information Scrambling in All-to-All Interacting Models},
  author        = {Saha, Abhik Kumar and Pathak, Tanay and Tezuka, Masaki},
  year          = {2026},
  eprint        = {2606.02207},
  archivePrefix = {arXiv},
  primaryClass  = {quant-ph},
  doi           = {10.48550/arXiv.2606.02207},
  url           = {https://arxiv.org/abs/2606.02207}
}

@article{9jbx-9kln,
  title = {Entanglement production in the Sachdev-Ye-Kitaev model and its variants},
  author = {Pathak, Tanay and Tezuka, Masaki},
  journal = {Phys. Rev. E},
  volume = {113},
  issue = {5},
  pages = {L052204},
  numpages = {7},
  year = {2026},
  month = {May},
  publisher = {American Physical Society},
  doi = {10.1103/9jbx-9kln},
  url = {https://link.aps.org/doi/10.1103/9jbx-9kln}
}

@article{PhysRevE.111.024201,
  title = {Relaxation fluctuations of correlation functions: Spin and random matrix models},
  author = {Pathak, Tanay},
  journal = {Phys. Rev. E},
  volume = {111},
  issue = {2},
  pages = {024201},
  numpages = {14},
  year = {2025},
  month = {Feb},
  publisher = {American Physical Society},
  doi = {10.1103/PhysRevE.111.024201},
  url = {https://link.aps.org/doi/10.1103/PhysRevE.111.024201}
}

\end{document}


\title{\textcolor{titleblue}{Supplemental Material:}
\textcolor{titleblue}{Full Eigenstate Thermalization in all-to-all Majorana Models}
}

\author{Abhik Kumar Saha\,\,\href{https://orcid.org/0000-0001-8168-6742}
{\includegraphics[scale=0.05]{orcidid.pdf}}}
\email{saha.abhikkumar.3k@kyoto-u.ac.jp}
\affiliation{Department of Physics, Kyoto University, Kitashirakawa Oiwakecho, Sakyo-ku, Kyoto 606-8502, Japan}

\maketitle
\raggedbottom
\onecolumngrid
\setcounter{figure}{0}
\renewcommand{\thefigure}{S\arabic{figure}}
\setcounter{figure}{0}
\renewcommand{\theequation}{S\arabic{equation}}
\setcounter{equation}{0}

In this Supplemental Material, we provide additional numerical results supporting the analysis of full eigenstate thermalization in chaotic and integrable all-to-all Majorana systems. In particular,
\begin{itemize}
    \item In Sec.~\ref{eq:supp_s1}, we examine the out-of-time-ordered correlator (OTOC) decomposition for the bilinear and quartic Majorana observables at smaller system sizes.
    \item In Sec.~\ref{eq:supp_s2}, we present time and frequency domain results for cactus factorization and crossing contributions in chaotic quartic SYK$_{4}$ Hamiltonian.
    \item In Sec.~\ref{eq:supp_s3}, we perform the corresponding analysis for integrable quadratic SYK$_{2}$ Hamiltonian.
    \item In Sec.~\ref{eq:supp_s4}, we investigate cactus factorization and crossing contribution in interacting disorder-free integrable quartic SYK$_{4}$ Hamiltonian.
\end{itemize}

\section{OTOC decomposition in Chaotic and integrable SYK$_{\rm q}$ models}\label{eq:supp_s1}

For a traceless Hermitian observable $\hat O$, the infinite-temperature OTOC is defined as $\mathrm{OTOC}(t)=\langle\hat O(t)\hat O(0)\hat O(t)\hat O(0)\rangle_{\beta=0}$. The approximate decomposition in terms of second and fourth-order ETH free cumulants reads \cite{PhysRevE.99.042139,PhysRevLett.129.170603,PhysRevLett.134.140404}
\begin{equation}
\mathrm{OTOC}(t)\simeq 2\bigl[k_2^{\mathrm{ETH}}(t)\bigr]^2+k_4^{\mathrm{ETH}}(t).
\label{eq:sm_otoc_decomposition}
\end{equation}
Here, $k_2^{\mathrm{ETH}}(t)$ and $k_4^{\mathrm{ETH}}(t)$ are defined through cyclic products of matrix elements with mutually distinct energy indices, as defined in the main text. The approximate relation in Eq.~\eqref{eq:sm_otoc_decomposition} relies on the suppression of crossing contributions and the factorization of cactus contributions.

\subsection{Chaotic quartic SYK$_{\rm 4}$}

\begin{figure}[h!]
    \centering
    \includegraphics[width=0.5\textwidth]{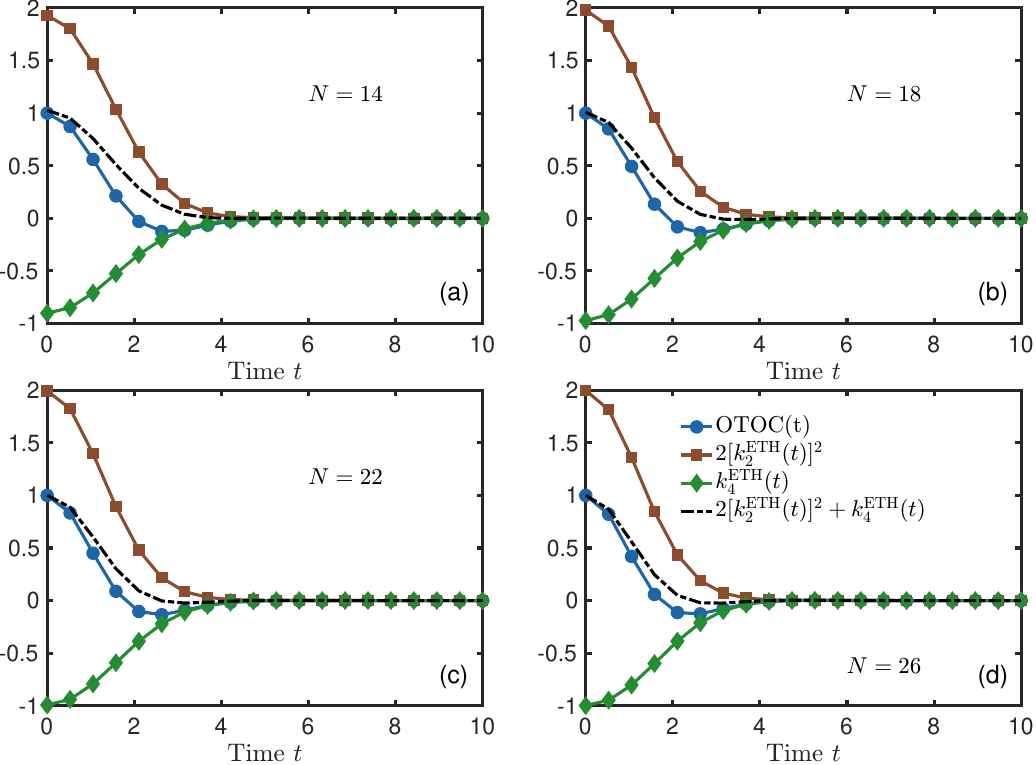}
    \caption{OTOC decomposition for the bilinear observable $\hat A=i\chi_1\chi_2$ in the chaotic quartic $\mathrm{SYK}_4$ model. The $\mathrm{OTOC}(t)$, the contributions $2[k_2^{\mathrm{ETH}}(t)]^2$ and $k_4^{\mathrm{ETH}}(t)$, and their sum are shown for (a) $N=14$, (b) $N=18$, (c) $N=22$, and (d) $N=26$. All results are obtained in the even-parity sector and averaged over $2^{20-N/2}$ independent disorder realizations for each system size.}
    \label{sfig:OTOC_SYK4_A}
\end{figure}

\begin{figure}[h]
    \centering
    \includegraphics[width=0.5\textwidth]{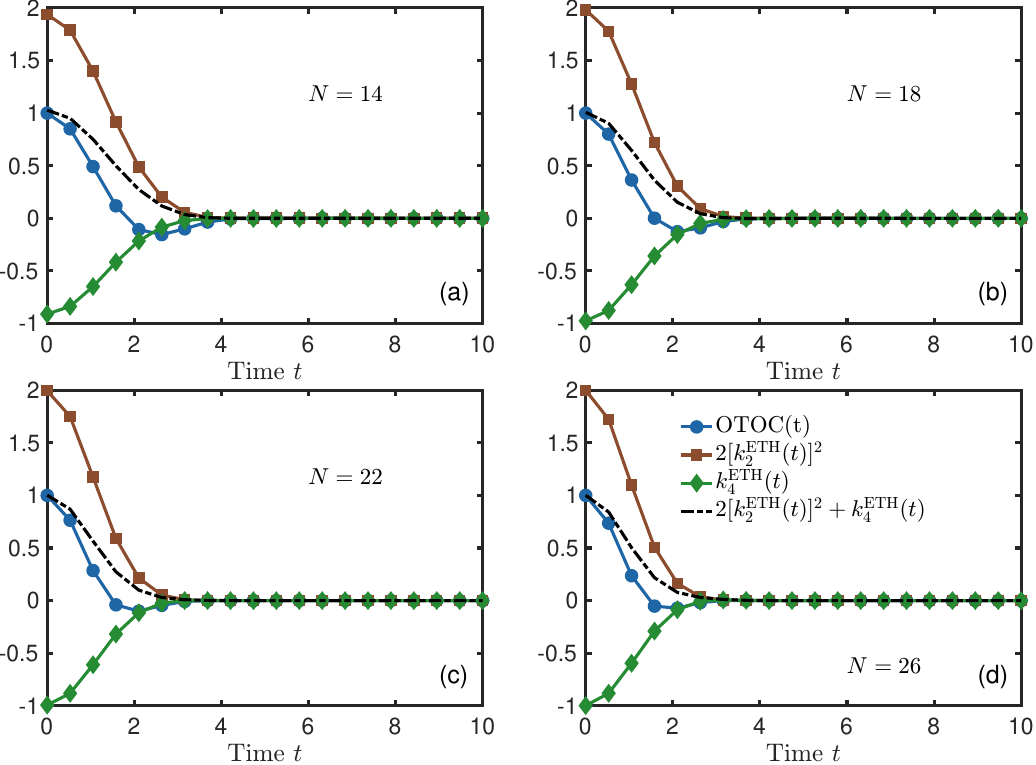}
    \caption{Same as Fig.~\ref{sfig:OTOC_SYK4_A}, but for the quartic observable $\hat B=\chi_1\chi_2\chi_3\chi_4$ in the chaotic quartic $\mathrm{SYK}_4$ model. Panels (a)-(d) correspond to $N=14,18,22,26$, respectively. All results are obtained in the even-parity sector and averaged over $2^{20-N/2}$ independent disorder realizations for each system size.}
    \label{sfig:OTOC_SYK4_B}
\end{figure}

We first consider the chaotic quartic $\mathrm{SYK}_4$ Hamiltonian. Figs.~\ref{sfig:OTOC_SYK4_A} and \ref{sfig:OTOC_SYK4_B} compare the $\mathrm{OTOC}(t)$ with $2[k_2^{\mathrm{ETH}}(t)]^2$, $k_4^{\mathrm{ETH}}(t)$, and their sum for $\hat A=i\chi_1\chi_2$ and $\hat B=\chi_1\chi_2\chi_3\chi_4$, respectively. For the bilinear observable $\hat A$ (Fig.~\ref{sfig:OTOC_SYK4_A}), the $\mathrm{OTOC}(t)$ decreases rapidly from its initial value, becomes slightly negative around $t\approx2$-$3$, and subsequently approaches zero. The second-order cumulant contribution, $2[k_2^{\rm ETH}(t)]^2$, decreases from a positive value, while the fourth-order cumulant $k_4^{\rm ETH}(t)$, increases from a negative value toward zero. Their sum approximately reproduces the overall $\mathrm{OTOC}(t)$ dynamics, although deviations are visible during the initial relaxation, particularly near the negative minimum. These deviations become smaller with increasing system size, consistent with finite-size corrections to the cumulant decomposition.

For the quartic observable (Fig.~\ref{sfig:OTOC_SYK4_B}), we observe similar relaxation behavior. The $\mathrm{OTOC}(t)$ rapidly decreases, reaches a small negative minimum, and subsequently approaches zero. The second-order cumulant contribution decreases from a positive value, whereas the fourth-order cumulant increases from a negative value toward zero. Their sum approximately reproduces the $\mathrm{OTOC}(t)$, with deviations becoming smaller as the system size increases, consistent with finite-size corrections to the cumulant decomposition. For both observables, the $\mathrm{OTOC}(t)$ and the cumulant contributions approach zero at $t\gg 4$. These results further support the approximate validity of the OTOC decomposition in chaotic all-to-all $\rm SYK_4$ Hamiltonian, consistent with the findings presented in the main text.

\subsection{Integrable quadratic SYK$_{\rm 2}$}
\begin{figure}[h!]
    \centering
    \includegraphics[width=0.5\textwidth]{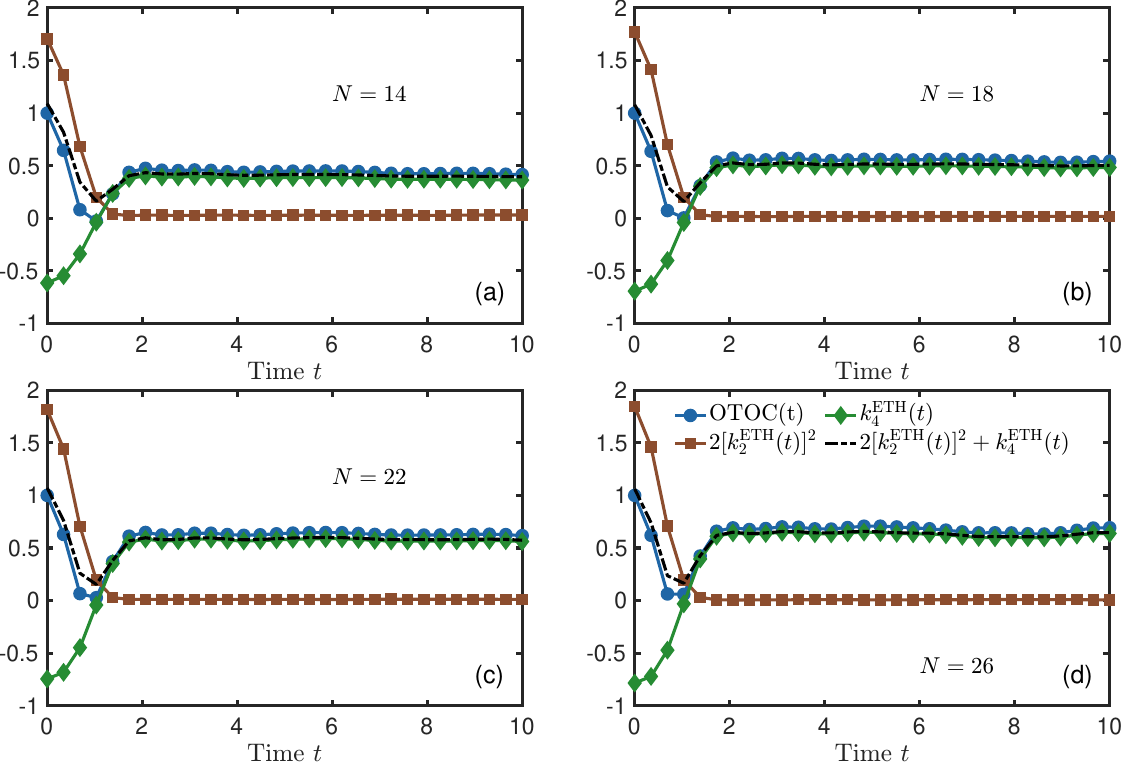}
    \caption{OTOC decomposition for the bilinear observable $\hat A=i\chi_1\chi_2$ in the integrable quadratic $\mathrm{SYK}_2$ model. The $\mathrm{OTOC}(t)$, the contributions $2[k_2^{\mathrm{ETH}}(t)]^2$ and $k_4^{\mathrm{ETH}}(t)$, and their sum are shown for (a) $N=14$, (b) $N=18$, (c) $N=22$, and (d) $N=26$. All results are obtained in the even-parity sector and averaged over $2^{20-N/2}$ independent disorder realizations for each system size.}
    \label{sfig:OTOC_SYK2_A}
\end{figure}

\begin{figure}[h]
    \centering
    \includegraphics[width=0.5\textwidth]{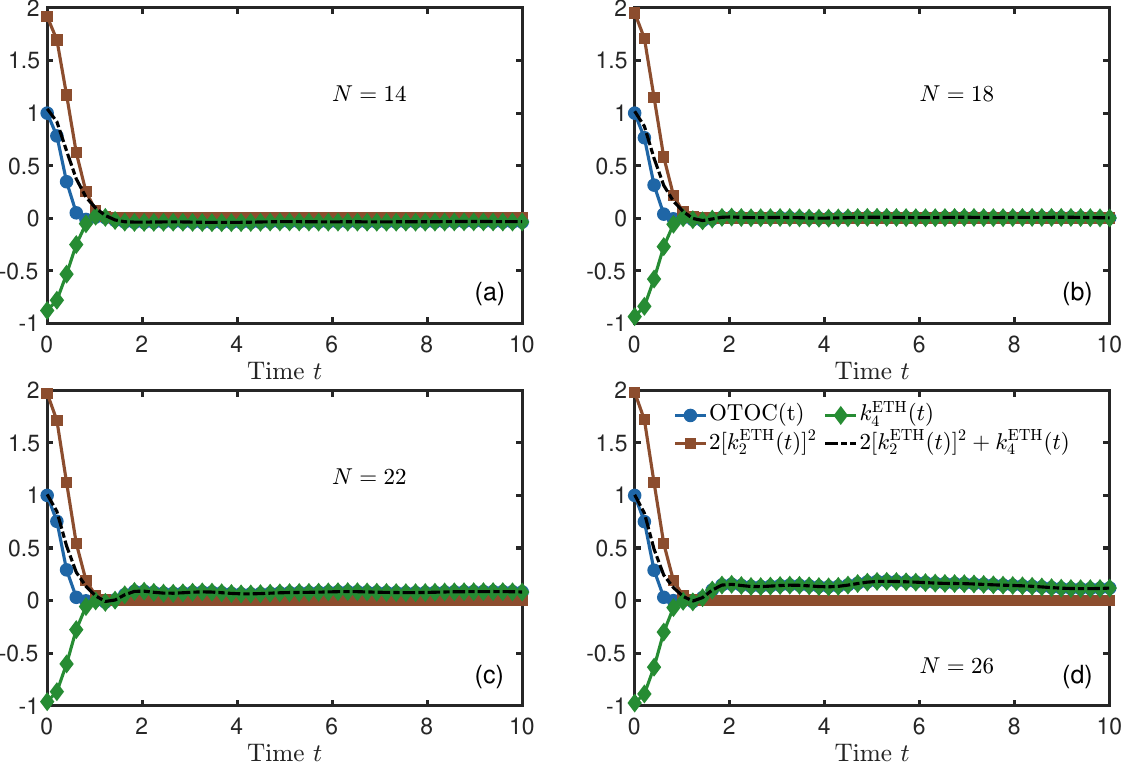}
    \caption{Same as Fig.~\ref{sfig:OTOC_SYK2_A}, but for the quartic observable $\hat B=\chi_1\chi_2\chi_3\chi_4$ in the integrable quadratic $\mathrm{SYK}_2$ model. Panels (a)-(d) correspond to $N=14,18,22,26$, respectively. All results are obtained in the even-parity sector and averaged over $2^{20-N/2}$ independent disorder realizations for each system size.}
    \label{sfig:OTOC_SYK2_B}
\end{figure}

We next consider the noninteracting integrable quadratic $\mathrm{SYK}_2$ model. Figs.~\ref{sfig:OTOC_SYK2_A} and \ref{sfig:OTOC_SYK2_B} show the time evolution of ${\rm OTOC}(t)$, the second and fourth-order ETH free cumulant contributions, and their sum for the bilinear observable $\hat A=i\chi_1\chi_2$ and the quartic observable $\hat B=\chi_1\chi_2\chi_3\chi_4$, respectively.

For the bilinear observable (Fig.~\ref{sfig:OTOC_SYK2_A}), the $\mathrm{OTOC}(t)$ initially decreases rapidly and reaches a value close to zero around $t\approx1$. Unlike chaotic SYK$_4$, it subsequently increases and approaches a finite positive value for $t\gg2$. This late-time value increases with system size over the range considered. The second-order contribution, $2[k_2^{\rm ETH}(t)]^2$, rapidly approaches zero, whereas the fourth-order cumulant, $k_4^{\rm ETH}(t)$, increases from a negative value and approaches a finite positive value. Consequently, the fourth-order cumulant dominates the late-time $\mathrm{OTOC}(t)$. Their sum approximately reproduces the initial decay and subsequent increase of the $\mathrm{OTOC}(t)$, with small deviations consistent with finite-size corrections to the cumulant decomposition.

For the quartic observable (Fig.~\ref{sfig:OTOC_SYK2_B}), the $\mathrm{OTOC}(t)$ also exhibits a rapid initial decay but displays a different late-time behavior. It remains close to zero for $N=14$ and $18$, while a small positive value develops for $N=22$ and $26$. The second-order contribution again rapidly approaches zero, whereas the fourth-order cumulant increases from a negative value and approaches a much smaller positive value than in the bilinear case. Their sum approximately reproduces the $\mathrm{OTOC}(t)$ dynamics, with deviations attributable to finite-size corrections. Thus, although the cumulant decomposition remains approximately valid for both observables, their late-time behavior differs considerably. In particular, $k_4^{\rm ETH}(t)$ captures the distinct relaxation dynamics and late-time $\mathrm{OTOC}(t)$ values of the bilinear and quartic observables in integrable SYK$_2$, in contrast to chaotic SYK$_4$, where the OTOCs for both observables approach zero.

\subsection{Interacting disorder-free integrable quartic SYK$_{\rm 4}$}
\begin{figure}[h!]
    \centering
    \includegraphics[width=0.5\textwidth]{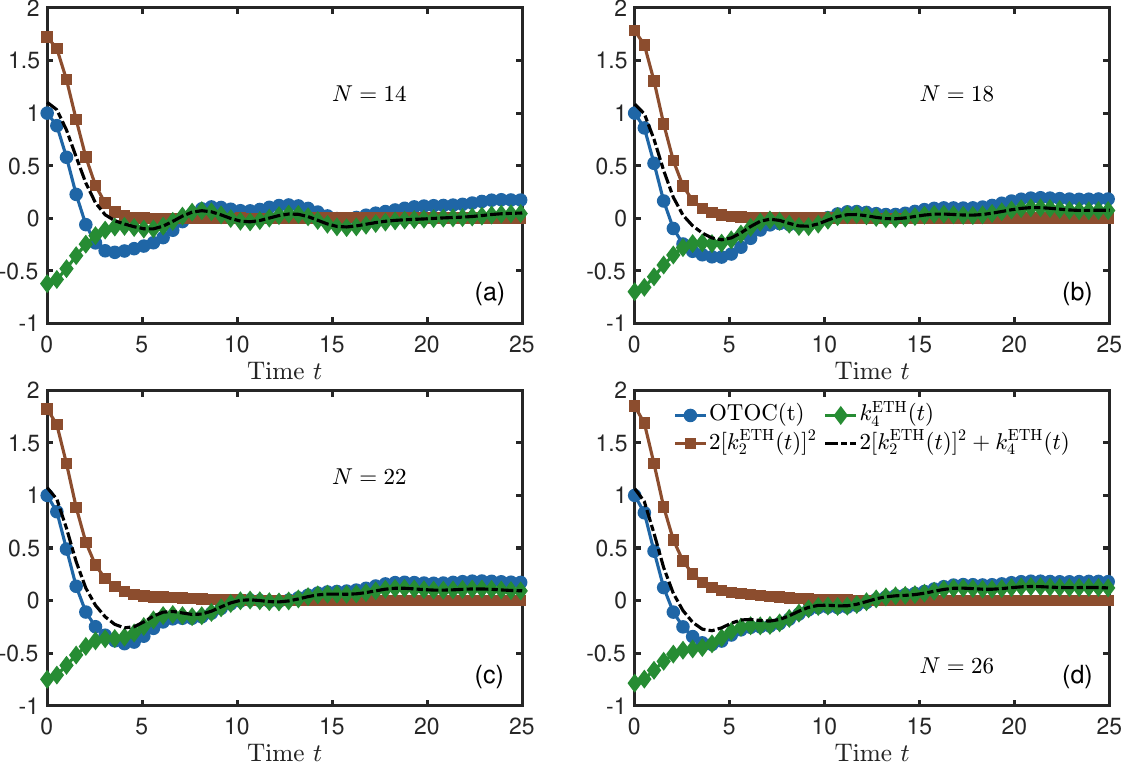}
    \caption{OTOC decomposition for the bilinear observable $\hat A=i\chi_1\chi_2$ in the interacting disorder-free integrable quartic $\mathrm{SYK}_4$ model. The $\mathrm{OTOC}(t)$, the contributions $2[k_2^{\mathrm{ETH}}(t)]^2$ and $k_4^{\mathrm{ETH}}(t)$, and their sum are shown for (a) $N=14$, (b) $N=18$, (c) $N=22$, and (d) $N=26$. All results are obtained in the even-parity sector.}
    \label{sfig:OTOC_SYK4_A_ND}
\end{figure}

\begin{figure}[h]
    \centering
    \includegraphics[width=0.5\textwidth]{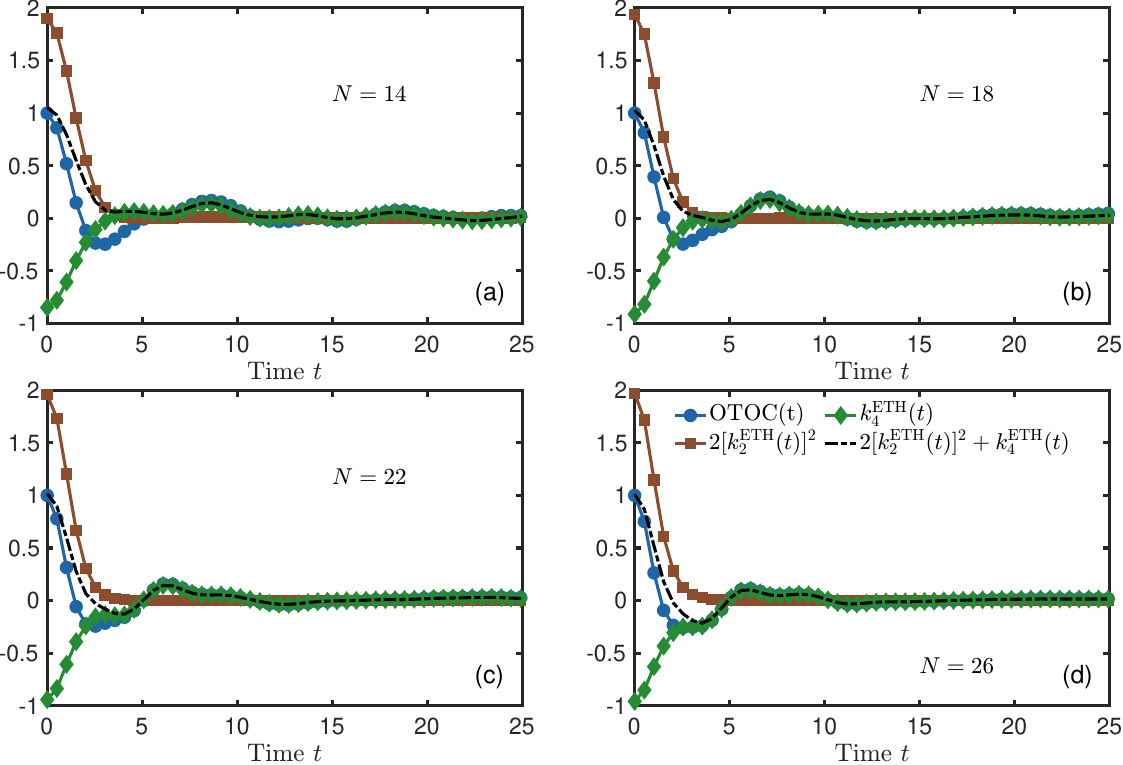}
    \caption{Same as Fig.~\ref{sfig:OTOC_SYK4_A_ND}, but for the quartic observable $\hat B=\chi_1\chi_2\chi_3\chi_4$ in the interacting disorder-free integrable quartic $\mathrm{SYK}_4$ model. Panels (a)-(d) correspond to $N=14,18,22,26$, respectively. All results are obtained in the even-parity sector.}
    \label{sfig:OTOC_SYK4_B_ND}
\end{figure}
We finally examine OTOC decomposition in the disorder-free integrable quartic $\mathrm{SYK}_4$ model with uniform four-body couplings. Figs.~\ref{sfig:OTOC_SYK4_A_ND} and \ref{sfig:OTOC_SYK4_B_ND} show the time evolution of ${\rm OTOC}(t)$, the second and fourth-order ETH free cumulant contributions, and their sum for the bilinear observable $\hat A=i\chi_1\chi_2$ and the quartic observable $\hat B=\chi_1\chi_2\chi_3\chi_4$, respectively.

For the bilinear observable (Fig.~\ref{sfig:OTOC_SYK4_A_ND}), the $\mathrm{OTOC}(t)$ initially decreases rapidly and reaches a negative minimum around $t\approx 3$-$5$. It subsequently increases toward a finite positive value, with nonmonotonic behavior that persists over a longer time interval. The second-order contribution, $2[k_2^{\rm ETH}(t)]^2$, decreases toward zero, while the fourth-order cumulant, $k_4^{\rm ETH}(t)$, increases from a negative value and gradually becomes positive. At later times, the fourth-order cumulant provides the dominant contribution to the $\mathrm{OTOC}(t)$. Their sum approximately reproduces the overall relaxation, although deviations are visible around the negative minimum, consistent with finite-size corrections to the cumulant decomposition.

For the quartic observable (Fig.~\ref{sfig:OTOC_SYK4_B_ND}), the $\mathrm{OTOC}(t)$ also initially decreases to a negative value but subsequently increases and exhibits a small positive maximum around $t\approx 7$-$8$. Unlike the bilinear case, it remains close to zero with small fluctuations at later times. The second-order contribution again decreases toward zero, whereas the fourth-order cumulant increases from a negative value and exhibits similar nonmonotonic behavior before approaching a small value. Their sum approximately captures the $\mathrm{OTOC}(t)$ dynamics for all system sizes, with visible finite-size deviations during the initial relaxation.

Thus, both observables exhibit nonmonotonic relaxation, but their late-time $\mathrm{OTOC}(t)$ values differ considerably, as reflected in the behavior of $k_4^{\rm ETH}(t)$. Compared with chaotic SYK$_4$, where both OTOCs rapidly approach zero, the disorder-free integrable model exhibits a longer relaxation process. Furthermore, unlike integrable SYK$_2$, where the bilinear $\mathrm{OTOC}(t)$ rapidly approaches a nearly stationary positive value, the corresponding relaxation in disorder-free integrable SYK$_4$ is more gradual and nonmonotonic. These results show that the approximate $\mathrm{OTOC}(t)$ cumulant decomposition remains valid in the interacting integrable model, while $k_4^{\rm ETH}(t)$ captures the differences in relaxation behavior between the two observables and between the two integrable Hamiltonians.

\section{Cactus factorization and crossing contributions in chaotic quartic SYK$_{\rm 4}$}\label{eq:supp_s2}
As discussed in the main text, the crossing contribution is defined as
\begin{equation}
{\rm cross}(t)=\frac{1}{\mathcal D}\sum_{i\neq j}
e^{i\omega_{ij}t}|O_{ij}|^4
=\mathcal O(\mathcal D^{-a}),\ a>0,
\end{equation}
while the cactus contribution is given by
\begin{equation}
\begin{aligned}
{\rm cac}(t_1,t_2)
&=\frac{1}{\mathcal D}
\sum_{i\neq j\neq k}
e^{i\omega_{ij}t_1+i\omega_{jk}t_2}
|O_{ij}|^2|O_{jk}|^2\\
&\approx k_2^{\rm ETH}(t_1)k_2^{\rm ETH}(t_2),
\end{aligned}
\label{seq:cactus}
\end{equation}
where $\omega_{ij}=E_i-E_j$ and $\mathcal D$ is the Hilbert-space dimension. Full ETH predicts that crossing contributions are exponentially subleading and cactus contributions factorize as $\mathrm{cac}(t_1,t_2)\approx k_2^{\rm ETH}(t_1)k_2^{\rm ETH}(t_2)$ in the thermodynamic limit. Following Ref.~\cite{PhysRevLett.134.140404}, we consider
\begin{equation}
{\rm cac}(t)={\rm cac}(t,t)
=\frac{1}{\mathcal D}\sum_{i\neq j\neq k}
e^{i(\omega_{ij}+\omega_{jk})t}
|O_{ij}|^2|O_{jk}|^2.
\end{equation}

\begin{figure}[h!]
    \centering
    \includegraphics[width=0.5\textwidth]{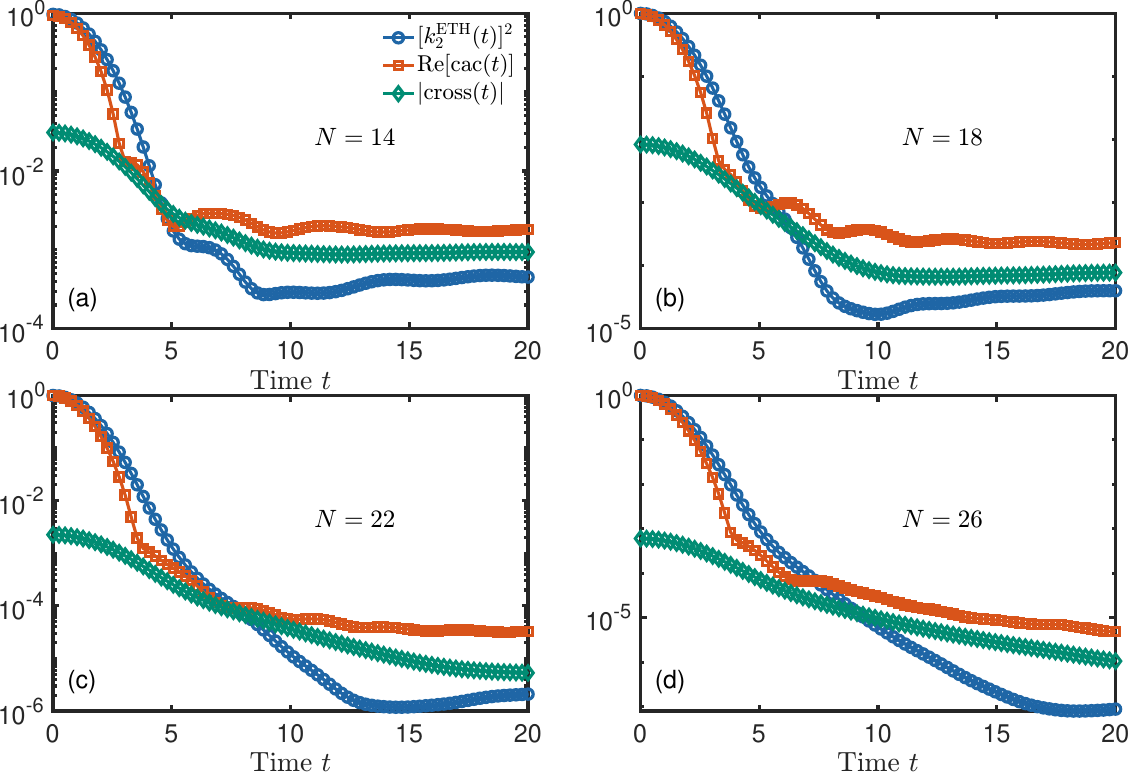}
    \caption{Time evolution of the cactus contribution $\mathrm{Re}[\mathrm{cac}(t)]$, the factorized second-order ETH free cumulant $[k_2^{\mathrm{ETH}}(t)]^2$, and the crossing contribution $|\mathrm{cross}(t)|$ for the bilinear observable $\hat A=i\chi_1\chi_2$ in the chaotic quartic $\mathrm{SYK}_4$ model. Results are shown for (a) $N=14$, (b) $N=18$, (c) $N=22$, and (d) $N=26$, with the vertical axis on a logarithmic scale. All results are obtained in the even-parity sector and averaged over $2^{20-N/2}$ independent disorder realizations for each system size.}
    \label{sfig:SYK4_cac_A}
\end{figure}
\begin{figure}[h!]
    \centering
    \includegraphics[width=0.5\textwidth]{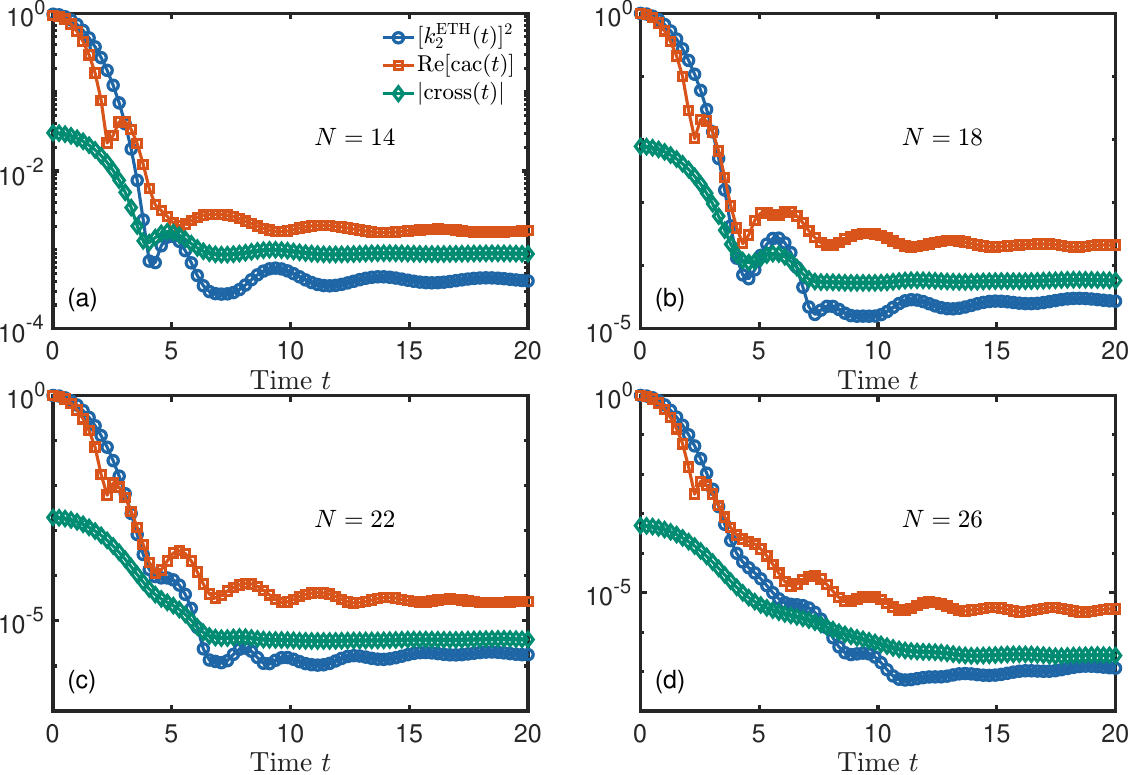}
    \caption{Same as Fig.~\ref{sfig:SYK4_cac_A}, but for the quartic observable $\hat B=\chi_1\chi_2\chi_3\chi_4$ in the chaotic quartic $\mathrm{SYK}_4$ model. Results are shown for (a) $N=14$, (b) $N=18$, (c) $N=22$, and (d) $N=26$. All results are obtained in the even-parity sector and averaged over $2^{20-N/2}$ independent disorder realizations for each system size.}
    \label{sfig:SYK4_cac_B}
\end{figure}

We now examine crossing contribution and cactus factorization in chaotic quartic SYK$_4$ for smaller system sizes, $N=14,18,22,$ and $26$, considering the bilinear observable $\hat A=i\chi_1\chi_2$ and the quartic observable $\hat B=\chi_1\chi_2\chi_3\chi_4$. Figs.~\ref{sfig:SYK4_cac_A} and \ref{sfig:SYK4_cac_B} show the time evolution of $\mathrm{Re}[\mathrm{cac}(t)]$, $[k_2^{\rm ETH}(t)]^2$, and $|\mathrm{cross}(t)|$ for the bilinear and quartic observables, respectively. For the bilinear observable (Fig.~\ref{sfig:SYK4_cac_A}), the cactus contribution approximately follows the factorized result during the initial decay. As time increases, deviations become visible, with the cactus contribution decreasing more slowly and approaching a small finite value, whereas the factorized contribution continues to decrease. The crossing contribution is considerably smaller during the initial relaxation, and its magnitude decreases with increasing system size.

For the quartic observable (Fig.~\ref{sfig:SYK4_cac_B}), we observe a similar initial decay of the cactus and factorized contributions, followed by a separation at later times. This separation becomes particularly visible at larger system sizes, where the factorized contribution decreases to much smaller values than the cactus contribution. The crossing contribution also becomes increasingly suppressed with system size. Thus, for both observables, cactus factorization approximately holds during the initial relaxation, while finite-size deviations remain visible at later times. These observations are consistent with the $N=30$ results presented in the main text.

We further investigate cactus factorization in the frequency domain, following Ref.~\cite{PhysRevLett.134.140404}. At infinite temperature, the frequency-resolved second-order ETH free cumulant is defined as
\begin{equation}
\widetilde{k}_2^{\rm ETH}(\omega)
=\frac{1}{\mathcal D}\sum_{i\neq j}
\delta_\tau(\omega-\omega_{ij})|O_{ij}|^2.
\end{equation}
The frequency-dependent cactus contribution corresponding to the definition in Eq.~\eqref{seq:cactus} is
\begin{equation}
\widetilde{\mathrm{cac}}(\omega_1,\omega_2)
=\frac{1}{\mathcal D}
\sum_{\substack{i\neq j \neq k}}
\delta_\tau(\omega_1-\omega_{ij})
\delta_\tau(\omega_2-\omega_{jk})
|O_{ij}|^2|O_{jk}|^2.
\end{equation}
Here, $\delta_\tau(\omega)$ denotes a smoothed delta function of width $1/\tau$, satisfying $\lim_{\tau\rightarrow\infty}\delta_\tau(\omega)=\delta(\omega)$. We choose a Gaussian function,
\begin{equation}
\delta_\tau(\omega)=
\frac{\tau}{\sqrt{2\pi}}e^{-\tau^2\omega^2/2},
\label{eq:sup_gaussian}
\end{equation}
with $\tau=3$ throughout our calculations. The full-ETH condition predicts the frequency-domain factorization
\begin{equation}
\widetilde{\mathrm{cac}}(\omega_1,\omega_2)
\approx
\widetilde{k}_2^{\rm ETH}(\omega_1)
\widetilde{k}_2^{\rm ETH}(\omega_2).
\label{eq:sup_freq_factorization}
\end{equation}
In the following, we examine this relation along the equal-frequency direction $\omega_1=\omega_2=\omega$.

\begin{figure}[h!]
    \centering
    \includegraphics[width=0.5\textwidth]{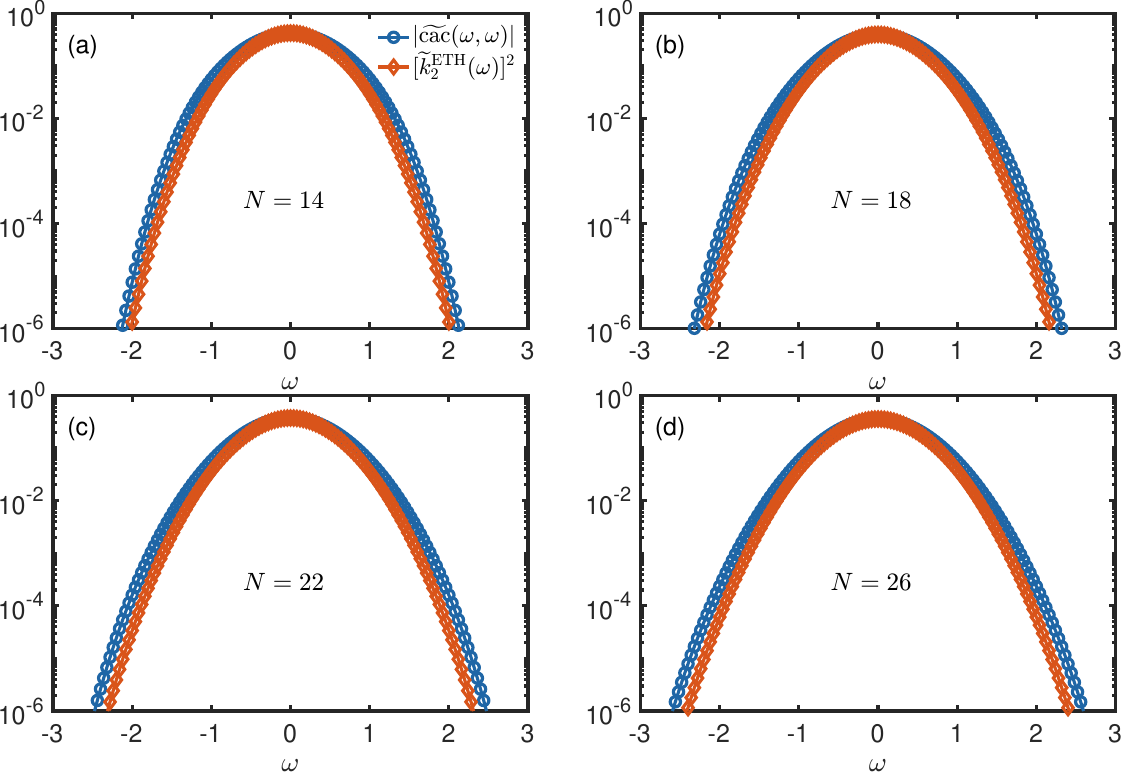}
    \caption{Frequency-domain comparison of the cactus contribution $|\widetilde{\mathrm{cac}}(\omega,\omega)|$ and the factorized second-order ETH free cumulant $[\widetilde{k}_2^{\mathrm{ETH}}(\omega)]^2$ along the equal-frequency direction $\omega_1=\omega_2=\omega$ for the bilinear observable $\hat A=i\chi_1\chi_2$ in the chaotic quartic $\mathrm{SYK}_4$ model. Results are shown for (a) $N=14$, (b) $N=18$, (c) $N=22$, and (d) $N=26$. All results are obtained in the even-parity sector and averaged over $2^{20-N/2}$ independent disorder realizations for each system size.}
    \label{sfig:SYK4_freq_A}
\end{figure}

\begin{figure}[h!]
    \centering
    \includegraphics[width=0.5\textwidth]{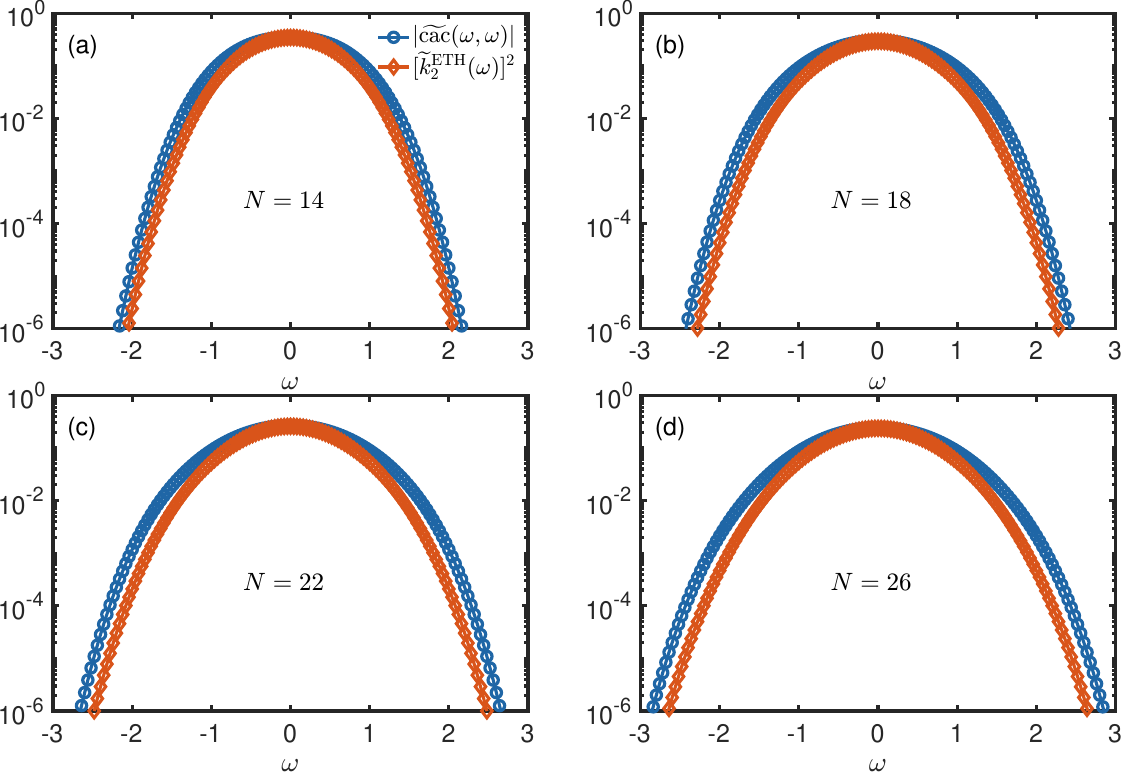}
    \caption{Same as Fig.~\ref{sfig:SYK4_freq_A}, but for the quartic observable $\hat B=\chi_1\chi_2\chi_3\chi_4$ in the chaotic quartic $\mathrm{SYK}_4$ model. Results are shown for (a) $N=14$, (b) $N=18$, (c) $N=22$, and (d) $N=26$. All results are obtained in the even-parity sector and averaged over $2^{20-N/2}$ independent disorder realizations for each system size.}
    \label{sfig:SYK4_freq_B}
\end{figure}

Figs.~\ref{sfig:SYK4_freq_A} and \ref{sfig:SYK4_freq_B} compare $|\widetilde{\mathrm{cac}}(\omega,\omega)|$ with $[\widetilde{k}_2^{\rm ETH}(\omega)]^2$ for the bilinear and quartic observables, respectively. For the bilinear observable (Fig.~\ref{sfig:SYK4_freq_A}), both quantities exhibit nearly symmetric frequency profiles, with a maximum near $\omega=0$ and a rapid decrease with increasing $|\omega|$. Although their frequency dependence is similar, the cactus contribution remains systematically larger than the factorized result for all system sizes considered.

For the quartic observable (Fig.~\ref{sfig:SYK4_freq_B}), the two contributions exhibit similar frequency profiles, but their separation is more pronounced than in the bilinear case. In particular, the cactus contribution remains above the factorized result over the displayed frequency range. These results show that the overall frequency dependence is approximately captured by the factorized second-order contribution, while finite-size deviations depend on the observable. Similar frequency-dependent behavior has been investigated in chaotic systems with local interactions~\cite{PhysRevLett.134.140404}. Our results demonstrate that the corresponding structure also appears in chaotic all-to-all SYK$_4$ Hamiltonian. We emphasize that the present calculations test cactus factorization along the equal-frequency direction, rather than over the full two-frequency plane.

\section{Cactus factorization and crossing contributions in integrable quadratic SYK$_{\rm 2}$}\label{eq:supp_s3}
\begin{figure}[h!]
    \centering
    \includegraphics[width=0.5\textwidth]{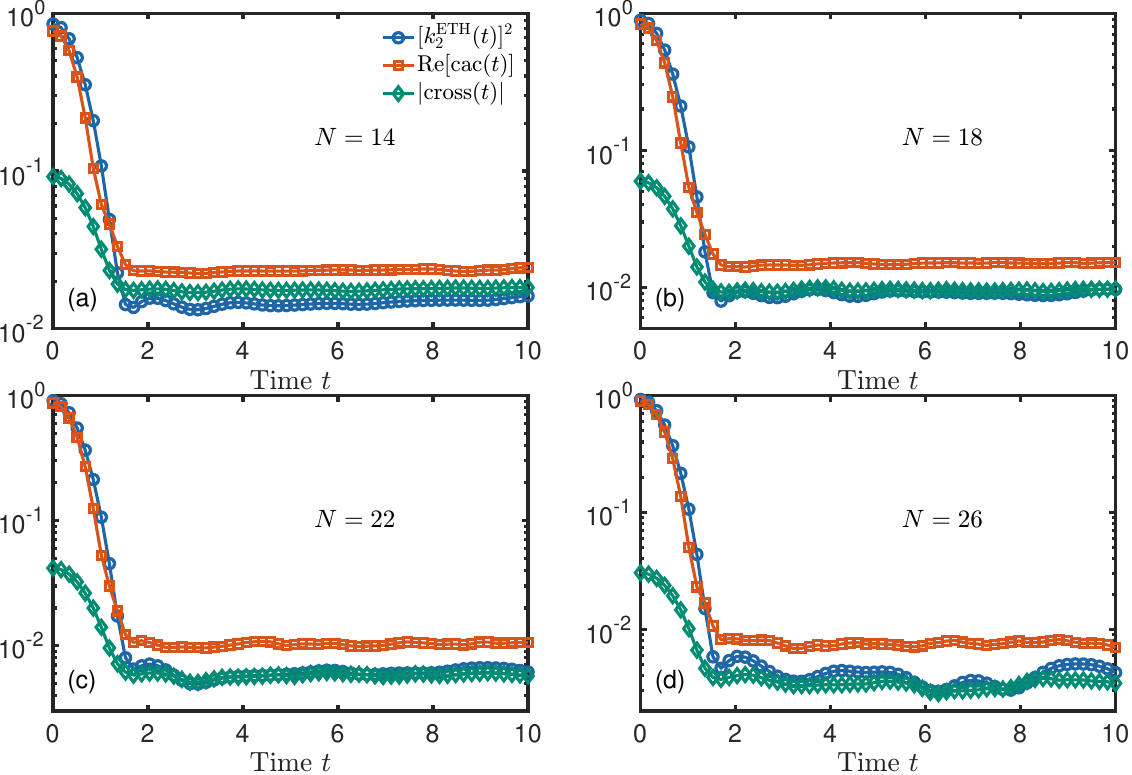}
    \caption{Time evolution of the cactus contribution $\mathrm{Re}[\mathrm{cac}(t)]$, the factorized second-order ETH free cumulant $[k_2^{\mathrm{ETH}}(t)]^2$, and the crossing contribution $|\mathrm{cross}(t)|$ for the bilinear observable $\hat A=i\chi_1\chi_2$ in the integrable quadratic $\mathrm{SYK}_2$ model. Results are shown for (a) $N=14$, (b) $N=18$, (c) $N=22$, and (d) $N=26$, with the vertical axis on a logarithmic scale. All results are obtained in the even-parity sector and averaged over $2^{20-N/2}$ independent disorder realizations for each system size.}
    \label{sfig:SYK2_cac_A}
\end{figure}

\begin{figure}[h!]
    \centering
    \includegraphics[width=0.5\textwidth]{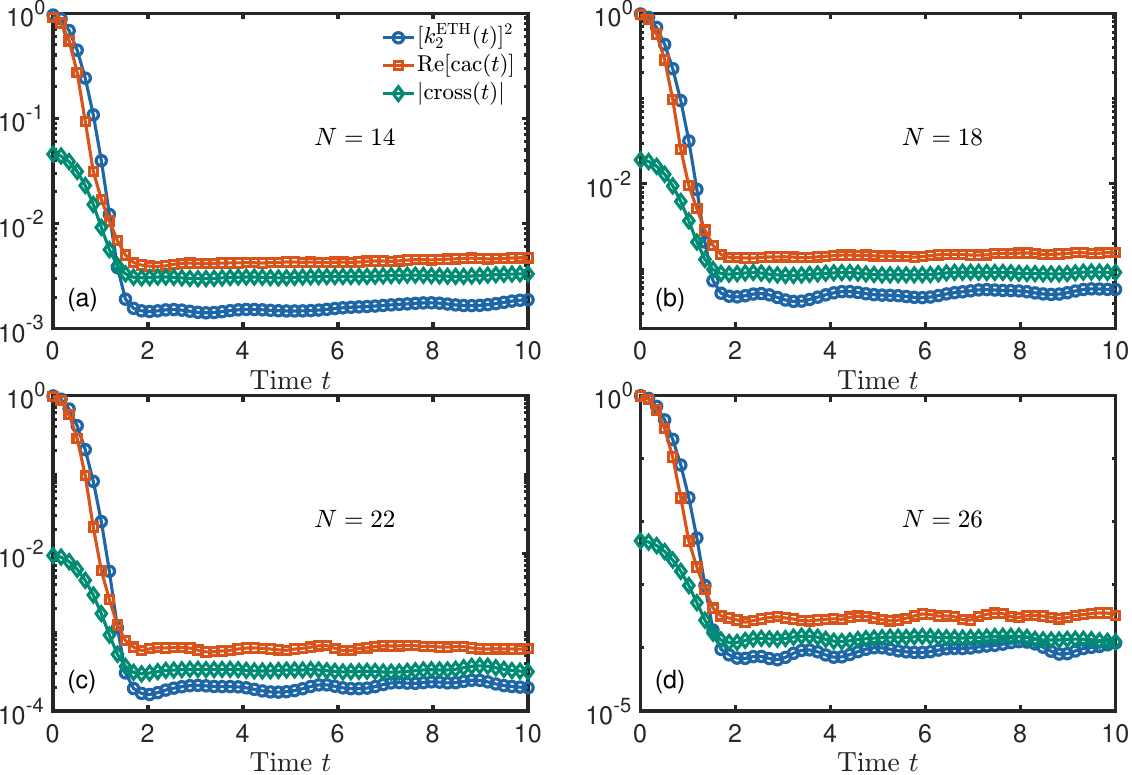}
    \caption{Same as Fig.~\ref{sfig:SYK2_cac_A}, but for the quartic observable $\hat B=\chi_1\chi_2\chi_3\chi_4$ in the integrable quadratic $\mathrm{SYK}_2$ model. Results are shown for (a) $N=14$, (b) $N=18$, (c) $N=22$, and (d) $N=26$. All results are obtained in the even-parity sector and averaged over $2^{20-N/2}$ independent disorder realizations for each system size.}
    \label{sfig:SYK2_cac_B}
\end{figure}

We next examine crossing contribution and cactus factorization in the noninteracting integrable quadratic SYK$_2$ model for smaller system sizes, $N=14,18,22,$ and $26$. Figs.~\ref{sfig:SYK2_cac_A} and \ref{sfig:SYK2_cac_B} show the time evolution of $\mathrm{Re}[\mathrm{cac}(t)]$, $[k_2^{\rm ETH}(t)]^2$, and $|\mathrm{cross}(t)|$ for the bilinear observable $\hat A=i\chi_1\chi_2$ and the quartic observable $\hat B=\chi_1\chi_2\chi_3\chi_4$, respectively.

For the bilinear observable (Fig.~\ref{sfig:SYK2_cac_A}), the cactus and factorized contributions initially decrease rapidly and approximately follow each other up to $t\approx1$. However, noticeable deviations develop around $t\approx2$, after which both contributions approach small finite values with persistent fluctuations. The cactus contribution remains larger than the factorized result, while the crossing contribution becomes comparable in magnitude at later times. Although all three contributions decrease with increasing system size, their qualitative behavior remains similar.

For the quartic observable (Fig.~\ref{sfig:SYK2_cac_B}), we observe a similar initial decay, followed by persistent fluctuations around finite values. However, the late-time magnitudes are considerably smaller than those for the bilinear observable, particularly at larger system sizes. The cactus contribution again remains above the factorized result, while the crossing contribution becomes comparable to both quantities. Unlike chaotic SYK$_4$, where the cactus and factorized contributions continue to decrease over a longer time interval, integrable SYK$_2$ exhibits a rapid initial decay followed by nearly stationary behavior. Moreover, the crossing contribution remains relatively significant at later times, consistent with the slower finite-size suppression discussed in the main text.

\begin{figure}[h!]
    \centering
    \includegraphics[width=0.5\textwidth]{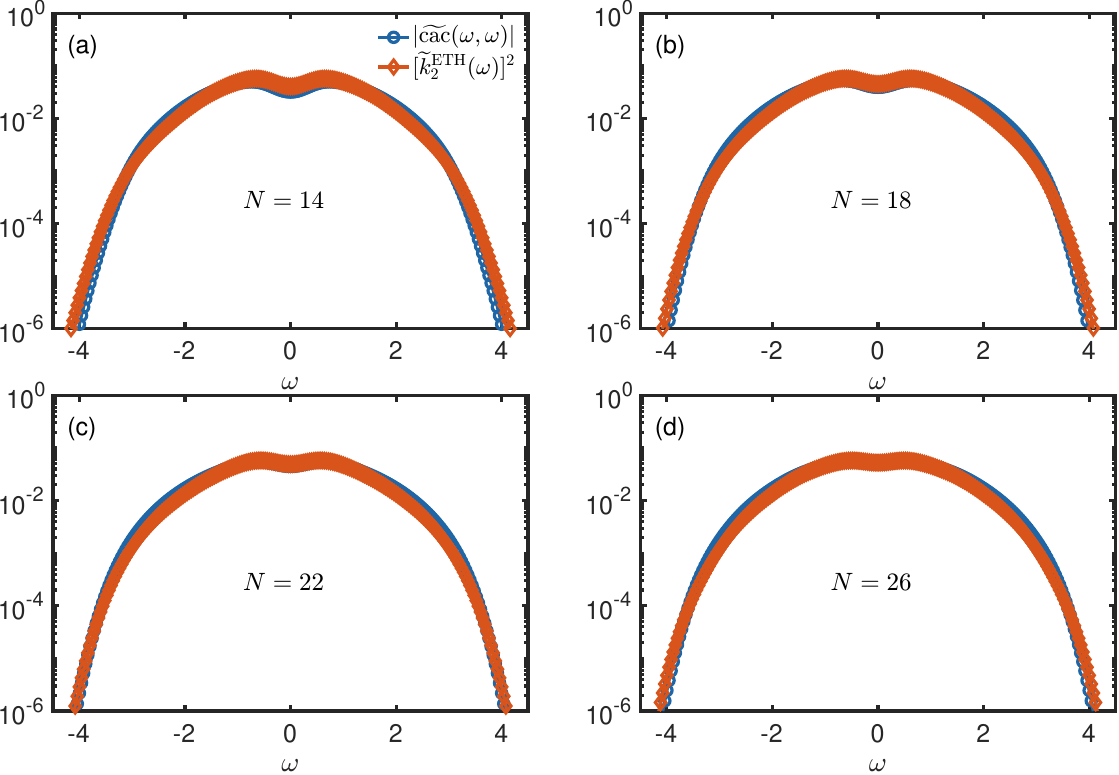}
    \caption{Frequency-domain comparison of the cactus contribution $|\widetilde{\mathrm{cac}}(\omega,\omega)|$ and the factorized second-order ETH free cumulant $[\widetilde{k}_2^{\mathrm{ETH}}(\omega)]^2$ along the equal-frequency direction $\omega_1=\omega_2=\omega$ for the bilinear observable $\hat A=i\chi_1\chi_2$ in the integrable quadratic $\mathrm{SYK}_2$ model. Results are shown for (a) $N=14$, (b) $N=18$, (c) $N=22$, and (d) $N=26$. All results are obtained in the even-parity sector and averaged over $2^{20-N/2}$ independent disorder realizations for each system size.}
    \label{sfig:SYK2_freq_A}
\end{figure}

\begin{figure}[h!]
    \centering
    \includegraphics[width=0.5\textwidth]{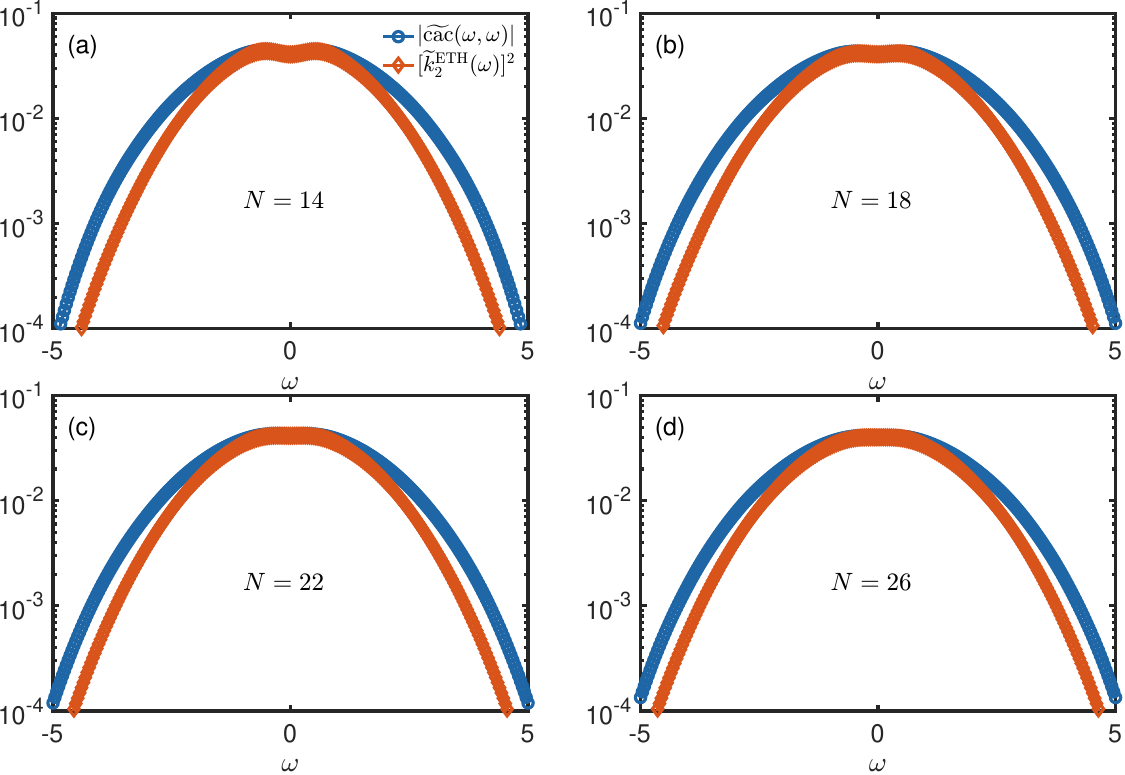}
    \caption{Same as Fig.~\ref{sfig:SYK2_freq_A}, but for the quartic observable $\hat B=\chi_1\chi_2\chi_3\chi_4$ in the integrable quadratic $\mathrm{SYK}_2$ model. Results are shown for (a) $N=14$, (b) $N=18$, (c) $N=22$, and (d) $N=26$. All results are obtained in the even-parity sector and averaged over $2^{20-N/2}$ independent disorder realizations for each system size.}
    \label{sfig:SYK2_freq_B}
\end{figure}

We further investigate cactus factorization in the frequency domain using the definitions introduced in the preceding section, with the same Gaussian broadening parameter $\tau=3$. Figs.~\ref{sfig:SYK2_freq_A} and \ref{sfig:SYK2_freq_B} compare $|\widetilde{\mathrm{cac}}(\omega,\omega)|$ with $[\widetilde{k}_2^{\rm ETH}(\omega)]^2$ for the bilinear and quartic observables, respectively. For the bilinear observable (Fig.~\ref{sfig:SYK2_freq_A}), both quantities exhibit nearly symmetric frequency profiles and approximately agree over a broad frequency range. However, small deviations are visible, particularly away from the central frequency region. Unlike chaotic SYK$_4$, where the frequency profiles exhibit a single maximum around $\omega=0$, the integrable SYK$_2$ model shows a shallow minimum near $\omega=0$, with two nearby maxima.

A similar frequency dependence is observed for the quartic observable (Fig.~\ref{sfig:SYK2_freq_B}), although the separation between the cactus and factorized contributions is more pronounced. The cactus contribution remains larger over most of the displayed frequency range, particularly at larger $|\omega|$. Both contributions nevertheless exhibit similar overall profiles for all system sizes considered. Thus, the frequency-domain results show approximate cactus factorization along the equal-frequency direction, with observable-dependent finite-size deviations. These results show that cactus factorization is approximately satisfied during the initial relaxation in integrable SYK$_2$, while noticeable deviations and relatively large crossing contributions persist at later times. In the frequency domain, the cactus and factorized contributions exhibit similar profiles, although their frequency dependence differs from that observed in chaotic SYK$_4$.

\section{Cactus factorization and crossing contributions in interacting disorder-free integrable quartic SYK$_{\rm 4}$}\label{eq:supp_s4}
\begin{figure}[h!]
    \centering
    \includegraphics[width=0.5\textwidth]{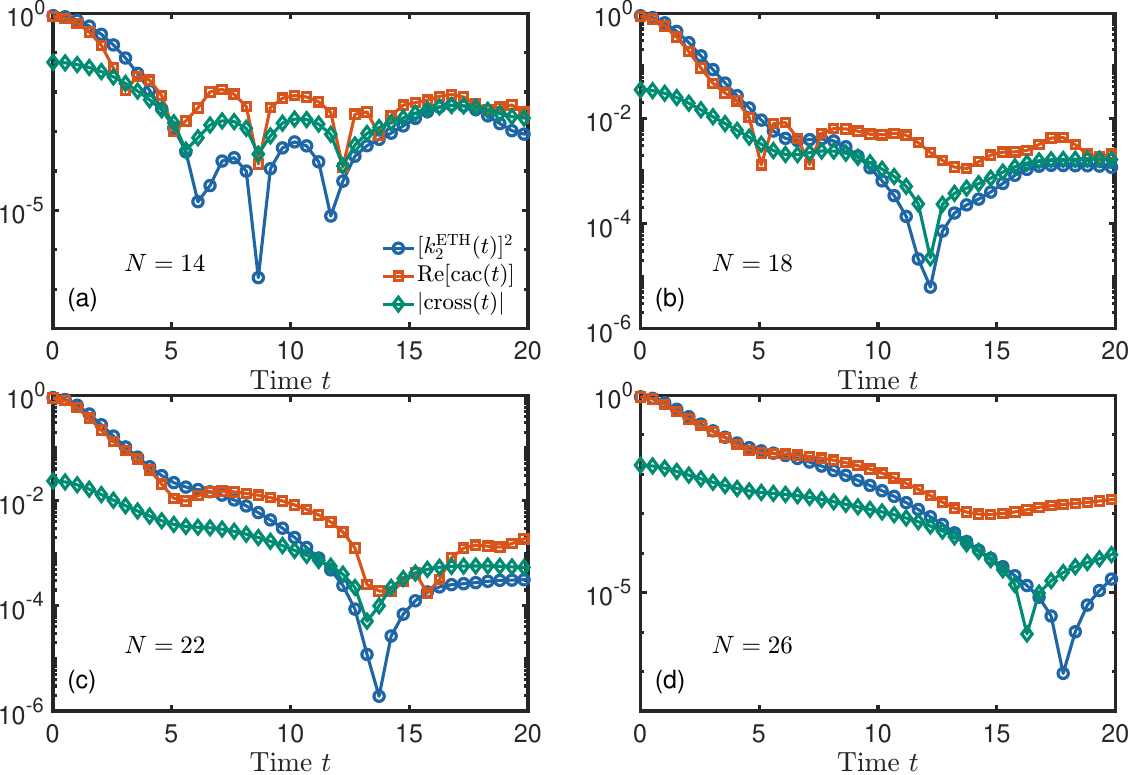}
    \caption{Time evolution of the cactus contribution $\mathrm{Re}[\mathrm{cac}(t)]$, the factorized second-order ETH free cumulant $[k_2^{\mathrm{ETH}}(t)]^2$, and the crossing contribution $|\mathrm{cross}(t)|$ for the bilinear observable $\hat A=i\chi_1\chi_2$ in the disorder-free integrable quartic $\mathrm{SYK}_4$ model. Results are shown for (a) $N=14$, (b) $N=18$, (c) $N=22$, and (d) $N=26$, with the vertical axis on a logarithmic scale. All results are obtained in the even-parity sector.}
    \label{sfig:SYK4ND_cac_A}
\end{figure}

\begin{figure}[h]
    \centering
    \includegraphics[width=0.5\textwidth]{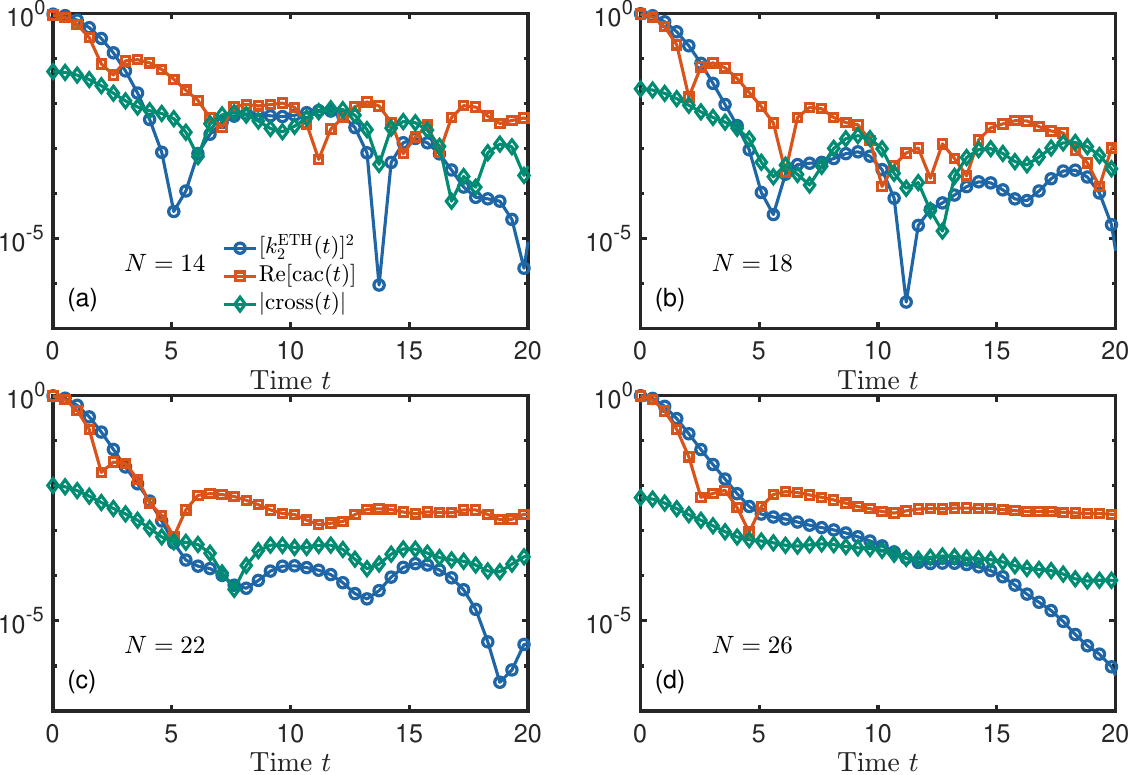}
    \caption{Same as Fig.~\ref{sfig:SYK4ND_cac_A}, but for the quartic observable $\hat B=\chi_1\chi_2\chi_3\chi_4$ in the disorder-free integrable quartic $\mathrm{SYK}_4$ model. Results are shown for (a) $N=14$, (b) $N=18$, (c) $N=22$, and (d) $N=26$. All results are obtained in the even-parity sector.}
    \label{sfig:SYK4ND_cac_B}
\end{figure}

Finally, we study the crossing contribution and cactus factorization in the interacting disorder-free integrable quartic $\mathrm{SYK}_4$ model for smaller system sizes, $N=14,18,22$, and $26$. Figs.~\ref{sfig:SYK4ND_cac_A} and \ref{sfig:SYK4ND_cac_B} show the time evolution of $\mathrm{Re}[\mathrm{cac}(t)]$, $[k_2^{\rm ETH}(t)]^2$, and $|\mathrm{cross}(t)|$ for the bilinear observable $\hat A=i\chi_1\chi_2$ and the quartic observable $\hat B=\chi_1\chi_2\chi_3\chi_4$, respectively.

For the bilinear observable (Fig.~\ref{sfig:SYK4ND_cac_A}), the cactus contribution approximately follows the factorized result during the initial decay, but noticeable deviations develop at later times. In particular, the cactus contribution decreases more slowly, while the factorized result continues to decrease and reaches much smaller values. For $N=14$ and $18$, both contributions exhibit pronounced fluctuations, whereas a smoother relaxation is observed for larger system sizes. The crossing contribution is initially smaller than the other two contributions but becomes comparable to the factorized result at later times. Unlike the noninteracting integrable SYK$_2$ model, where the contributions rapidly decrease and approach nearly stationary values around $t\approx2$, the disorder-free integrable SYK$_4$ model exhibits a much slower and nonmonotonic relaxation.

For the quartic observable (Fig.~\ref{sfig:SYK4ND_cac_B}), we observe similar behavior, although deviations from cactus factorization become visible earlier during the relaxation, particularly for larger system sizes. The cactus contribution remains larger than the factorized result at later times, while the crossing contribution becomes comparable to the other contributions in parts of the time window. The fluctuations are more pronounced for smaller system sizes and become less prominent as $N$ increases. Compared with chaotic SYK$_4$, where the cactus and factorized contributions decay rapidly, the disorder-free integrable SYK$_4$ model exhibits prolonged relaxation and persistent deviations from factorization. These observations are consistent with the $N=30$ results presented in the main text.

\begin{figure}[h!]
    \centering
    \includegraphics[width=0.5\textwidth]{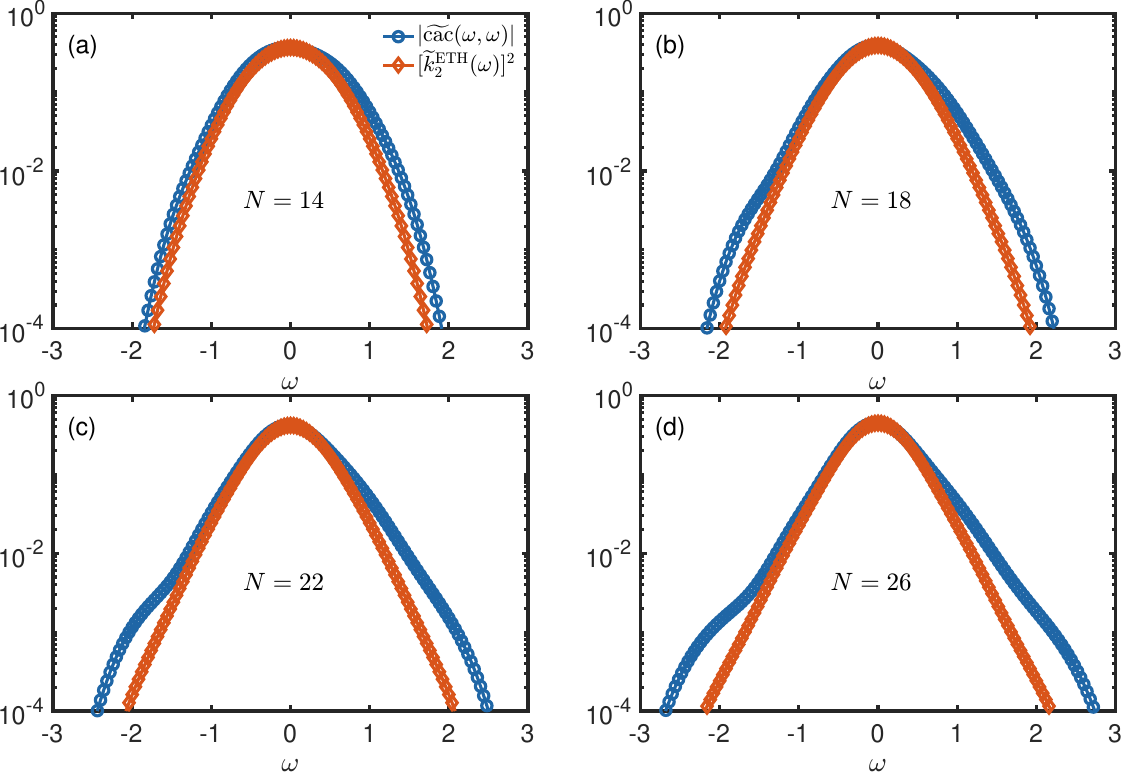}
    \caption{Frequency-domain comparison of the cactus contribution $|\widetilde{\mathrm{cac}}(\omega,\omega)|$ and the factorized second-order ETH free cumulant $[\widetilde{k}_2^{\mathrm{ETH}}(\omega)]^2$ along the equal-frequency direction $\omega_1=\omega_2=\omega$ for the bilinear observable $\hat A=i\chi_1\chi_2$ in the disorder-free integrable quartic $\mathrm{SYK}_4$ model. Results are shown for (a) $N=14$, (b) $N=18$, (c) $N=22$, and (d) $N=26$. All results are obtained in the even-parity sector.}
    \label{sfig:SYK4ND_freq_A}
\end{figure}

\begin{figure}[h]
    \centering
    \includegraphics[width=0.5\textwidth]{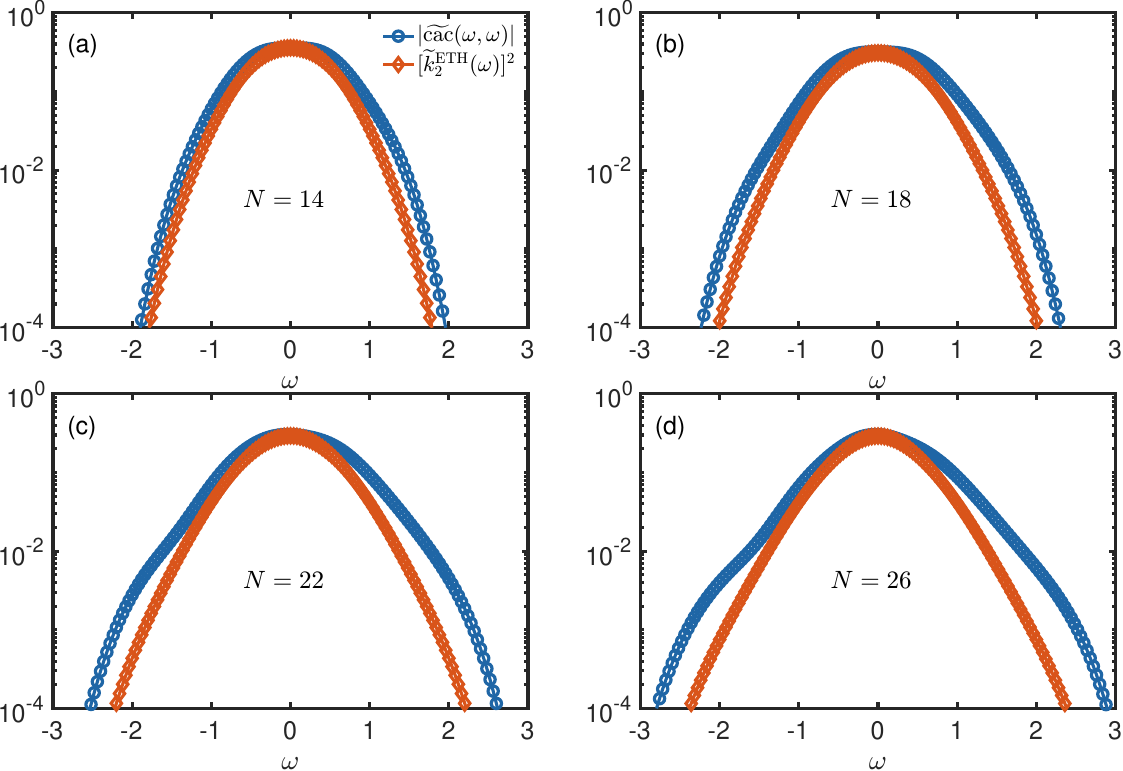}
    \caption{Same as Fig.~\ref{sfig:SYK4ND_freq_A}, but for the quartic observable $\hat B=\chi_1\chi_2\chi_3\chi_4$ in the disorder-free integrable quartic $\mathrm{SYK}_4$ model. Results are shown for (a) $N=14$, (b) $N=18$, (c) $N=22$, and (d) $N=26$. All results are obtained in the even-parity sector.}
    \label{sfig:SYK4ND_freq_B}
\end{figure}

We further investigate cactus factorization in the frequency domain using the definitions introduced previously, with Gaussian broadening $\tau=3$. Figs.~\ref{sfig:SYK4ND_freq_A} and \ref{sfig:SYK4ND_freq_B} compare $|\widetilde{\mathrm{cac}}(\omega,\omega)|$ with $[\widetilde{k}_2^{\rm ETH}(\omega)]^2$ for the bilinear and quartic observables, respectively. For the bilinear observable (Fig.~\ref{sfig:SYK4ND_freq_A}), both quantities exhibit nearly symmetric frequency profiles with a maximum around $\omega=0$ and decrease with increasing $|\omega|$. Although their overall frequency dependence is similar, the cactus contribution remains larger than the factorized result, with more noticeable deviations away from $\omega=0$, particularly for $N=22$ and $26$.

Similar behavior is observed for the quartic observable (Fig.~\ref{sfig:SYK4ND_freq_B}), where the two contributions exhibit comparable frequency profiles but remain separated over the displayed frequency range. The difference becomes particularly visible at larger $|\omega|$ for the larger system sizes. Interestingly, these single-peaked frequency profiles resemble those observed in chaotic SYK$_4$, whereas integrable SYK$_2$ exhibits a shallow minimum near $\omega=0$ with two nearby maxima. Thus, despite their different relaxation dynamics, chaotic and disorder-free integrable SYK$_4$ exhibit similar qualitative frequency dependence along the equal-frequency direction. However, noticeable deviations from cactus factorization persist in the disorder-free model. Together, these results show that the time dependence and the finite-size corrections to cactus factorization depend on the interaction structure, even among all-to-all Majorana systems.

\bibliography{references}